\documentclass[aps,prx,letter,preprintnumbers,superscriptaddress,nofootinbib,longbibliography,floatfix]{revtex4-2}
\pdfoutput=1
\usepackage{rotating}
\usepackage{array}
\usepackage{amsmath}
\usepackage[normalem]{ulem}
\usepackage{slashed}
\usepackage{booktabs}
\usepackage[pdftex,table]{xcolor}
\usepackage{units}
\usepackage{xfrac}
\usepackage{mathtools}
\usepackage{empheq}
\usepackage[]{units}
\usepackage{multirow}
\usepackage{amssymb}
\usepackage{url}
\usepackage{comment}
\usepackage{physics}
\usepackage{color,soul}
\usepackage{bbm}
\usepackage{adjustbox}
\usepackage[T1]{fontenc}
\usepackage{xspace}

\usepackage{graphicx,caption,subcaption,tikz}
\usetikzlibrary{backgrounds}
\setcitestyle{authoryear,round}
\usetikzlibrary{calc}
\DeclareCaptionFormat{overlay}{\gdef\capoverlay{#1#2#3\par}}
\DeclareCaptionStyle{overlay}{format=overlay}
\DeclareCaptionFormat{suboverlay}{\gdef\subcapoverlay{(\thesubfigure) #3\par}}
\DeclareCaptionStyle{suboverlay}{format=suboverlay}

\usepackage{hyperref}
\hypersetup{
  colorlinks=true,
  citecolor=blue,
  linkcolor=blue,
  urlcolor=blue
}
\makeatletter
\renewcommand{\paragraph}{%
  \@startsection{paragraph}{4}{\z@}%
  {1.5ex \@plus 0.5ex \@minus 0.2ex}
  {-1em}
  {\normalfont\normalsize\bfseries}
}
\makeatother

\newcommand{\vpm}[2]{#1\,{\scriptstyle\pm#2}}
\newcommand{\pf}{\textit{PF}\xspace}
\newcommand{\pflow}{\textit{PFlow}\xspace}
\renewcommand{\tr}{\textit{TR}\xspace}
\newcommand{\truth}{\textit{Truth}\xspace}
\newcommand{\fs}{\textit{FS}\xspace}
\newcommand{\fastsim}{\textit{FastSim}\xspace}

\begin{document}

\title{Scaling Collider Event Generation with Residual-Quantized Tokens}
\author{Dan Godi}
\email{dan.godi@weizmann.ac.il}
\author{Dmitrii Kobylianskii}
\author{Eilam Gross}
\affiliation{Weizmann Institute of Science, Rehovot, Israel}
\title{Scaling Collider Event Generation with Residual-Quantized Tokens}

\begin{abstract}
\noindent Full detector simulation and reconstruction of collider events are projected to become major bottlenecks at the High-Luminosity Large Hadron Collider, motivating the development of fast, ML-based surrogates. At the same time, LLMs have driven fast progress in generative discrete modeling: autoregressive transformers trained on tokenized data now represent the state of the art across a range of generative tasks. We extend the discrete modeling paradigm by introducing a particle-level generative model trained on residual-quantized full-event data. We demonstrate the ability of this model family to perform conditional generation from detector-stable particles; we study its scaling behavior across a range of dataset and model sizes, characterize the effects of repeated data exposure and demonstrate that token-level loss systematically predicts downstream physical fidelity. These results provide an empirical framework for scalable collider full-event generation based on residual-quantized representations.
 
\end{abstract}

\maketitle

\section{Introduction}
\label{sec:introduction}
At collider experiments, the growing sample size requirements for synthetic datasets are set to become a significant computational bottleneck. Projections for the High-Luminosity upgrade of the Large Hadron Collider (HL-LHC) estimate that, under current paradigms, full simulation tools will outstrip available compute resources, such that simulation and reconstruction of synthetic datasets will exceed the processing requirements of real data~\citep{CERN-LHCC-2022-005, Collaboration:2957472}. Fast Simulation (\fastsim) methods have emerged as solutions to this issue; these techniques use approximations to avoid modeling the full physical aspects of the simulation, hence consuming significantly fewer resources.

Over the past several years, Machine Learning (ML) has become a central tool in High Energy Physics (HEP), with deep learning algorithms now seeing wide adoption across the analysis chain~\citep{butterMachineLearningLHC2023a, krauseCaloChallenge2022Community2025, aadTransformingJetFlavour2026}.
Concurrently, advances in discrete sequence generation have established robust architectures and well-characterized scaling laws. These developments have been adopted in scientific applications, motivating recent formulations of HEP simulation as a discrete modeling problem.

\paragraph{Related work}
Previous work in the field has focused on modeling at the jet~\citep{birkOmniJet$a$FirstCrosstask2024, gollingMaskedParticleModeling2024b, amramAspenOpenJets2024, bhimjiOmniLearnedFoundationModel2026} and calorimeter~\citep{birkOmniJet$a_C$LearningPoint2025, cardona-giraldoGeneralizableFoundationModels2026} levels. More recently, generative approaches have been extended to conditional full-event simulation~\citep{dreyerConditionalDeepGenerative2026}, but focusing on flow-matching techniques. 
Separately, neural scaling laws have been investigated in the context of jet tagging and generation~\citep{vigl2026neuralscalinglawsboosted,amram2026neuralscalinglawsjet}, building on broader developments in model and data scaling. In parallel with our work,~\citet{kakati2026promptingparticlephysicstokenized} have explored a multimodal generative approach based on residual quantization in the single-jet setting.

In this work, we extend the discrete modeling paradigm to the full-event level.
Full-event simulation is typically a two-stage process. Simulating the initial collision, subsequent interactions, and particle decays is relatively inexpensive, whereas simulating the detector response to stable particles and reconstructing the resulting signals is substantially more computationally intensive.
The final output particles that are obtained after reconstruction are referred to as Particle Flow (\pf) objects; these objects approximate the original set of particles entering the detector (Truth particles, \tr), and are expected to maintain the physical properties and metrics used for downstream analysis. 
In principle, ML-based fast-simulation algorithms should be able to generate \pf events without the need for \tr conditioning, but conditional generation (and thus modeling of the detector response instead of the final output distribution) is strongly preferred for the ability to generalize to physical processes (rare or novel/theoretical) not seen during training. 
Since the model is conditioned on \tr-level input, it learns the detector response itself as a physical map that is in principle independent of the process that produced the incoming particles, so a single surrogate can serve many signals and backgrounds rather than being re-tuned per analysis. 
While one could directly generate high-level observables, such as jets (sprays of particles clustered together), producing \pf objects directly preserves the full event structure. This ensures compatibility with the existing downstream pipelines while providing a flexible foundation for developing new algorithms.

\paragraph{Contribution}
Building on this full-event formulation, we show how the benefits of discrete residual-quantized representations transfer to point-cloud generation.
We then study the scaling behavior of models spanning 10M to 1B parameters, characterize their dependence on model and dataset size, and examine the effects of repeated data exposure.
We further demonstrate that the achieved loss is a powerful predictor of modeling performance at the physical scale: improvements in cross-entropy translate predictably into higher fidelity across particle-, jet-, and global event observables.
These findings demonstrate that token-level scaling laws can reliably guide model capacity, dataset size, and compute allocation when designing generative surrogates for collider physics.

\section{Methods}
\label{sec:methods}
Our approach factorizes event generation into two stages. First, a quantizer network is trained to map each truth or reconstructed particle to a short sequence of discrete tokens drawn from a learned codebook, resulting in a fixed vocabulary in which an event is represented as a sequence of integers. 
Second, an autoregressive transformer models the distribution over these token sequences, so that generation reduces to sampling the next token from a categorical distribution.
Once a token sequence is sampled, it is passed back through the autoencoder's decoder to recover continuous particle-level quantities.
Decoupling the generative model from the data representation allows the transformer to operate on a compact discrete space, while the mapping to and from physical observables is handled by the separately trained fixed tokenizer.

\subsection{Dataset}
The model is trained and tested on several physical processes, from the CMS Open Data 2011 collection.
Events are generated with Pythia 6.4.25~\citep{sjostrandBriefIntroductionPYTHIA2008}, passed through Geant4~\citep{agostinelliGeant4aSimulationToolkit2003} simulation of the CMS detector, and subsequently reconstructed using the CMS Particle Flow algorithm~\citep{CMS-PAS-PFT-09-001}.

In each event we select from the dataset the set of generator-level final state \tr particles and the corresponding set of \pf particles. 
Each particle from these collections is described by its transverse momentum $p_T$, its pseudorapidity $\eta$ and azimuthal angle $\phi$, its class and the coordinates of its production vertex. 
Particles are categorized into one of five classes: charged hadron, electron, muon, neutral hadron, photon, and neutral \pf particles have zeroed vertices (in real datasets the vertex of neutral particles can never be reconstructed since the tracking system only reads charged-particle deposits).
Finally, the particles in each event are $p_T$-ordered, from hardest to softest.
\\To improve momentum reconstruction and lessen tracker inefficiencies, we follow \citet{dreyerConditionalDeepGenerative2026} and require $p_T>1$ GeV for \pf particles, and apply $p_T > 0.25$ GeV cut on the \tr set to reduce its cardinality.

\begin{table}[ht]
\caption{Datasets used for training and testing, in-distribution testing slices sum up to $50,000$ events. Parton-level $p_T$ ranges are shown for QCD samples.}
\label{tab:datasets}
\centering

\renewcommand{\arraystretch}{1.15}
\setlength{\tabcolsep}{7pt}

\begin{tabular}{lcccc}
\multicolumn{1}{c}{\bf Dataset} &
\multicolumn{1}{c}{\bf $p_{\mathrm{T}}^{\min}$--$p_{\mathrm{T}}^{\max}$ [GeV]} &
\multicolumn{1}{c}{\bf Type} &
\multicolumn{1}{c}{\bf Training [evts]} &
\multicolumn{1}{c}{\bf Testing [evts]}
\\ \hline \\[-6pt]

$H \rightarrow 4\ell$ &
-- &
Out-of-distribution &
&
$50{,}000$
\\

$t\bar{t}$ &
-- &
In distribution &
$9{,}361{,}338$ &
$28{,}263$
\\

QCD  &
470--600 &
In distribution &
$3{,}737{,}811$ &
$11{,}285$
\\

QCD  &
600--800 &
In distribution &
$3{,}462{,}168$ &
$10{,}452$
\\

QCD  &
1000--1400 &
Out-of-distribution &
&
$50{,}000$
\\

\end{tabular}
\end{table}

The training dataset is made of a mixture of $t\bar{t}$ and QCD samples, available with high statistics. The model is subsequently tested on a mixture of statistically identical held-out samples, and a series of out-of-distribution samples of physical processes not seen during training \citep{h4lep_data,qcd1000_data,qcd470_data,qcd600_data,ttbar_data}. An overview is available in Table~\ref{tab:datasets}.

\subsection{RQ-VAE}
\label{sec:rqvae}
Tokenizer networks such as vector-quantized autoencoders \citep{oord_neural_2018} (VQ-VAEs) compress an input into a discrete latent vector (code). An encoder first maps an input $x$ to a continuous latent $z$, which is then quantized against a learned codebook $\mathcal{C} = \{e_k\}^K_{k=1}$ by nearest-neighbor lookup, so that $x$ is represented by a single index $k = \mathcal{Q}(z;\mathcal{C})$ and decoded from $e_k$:
\begin{equation}
  \mathcal{Q}(z;\mathcal{C}) = \mathrm{argmin}_{k \in \{1,\dots,K\}}
  ||z-e_k||_2 ,
  \label{eq:vq}
\end{equation}
The fidelity of this representation is limited by the codebook size $K$: driving down the quantization error requires a finer partition of the latent space. In practice, large codebooks are difficult to train and suffer from codebook collapse, in which a small fraction of entries are ever selected and the effective capacity saturates below $K$.

By representing $z$ with an ordered sequence of $M$ codes, residual quantization~\citep{zeghidourSoundStreamEndtoEndNeural2021} avoids this tradeoff. The effective codebook size increases to $K^{M}$ while avoiding having to scale $K$ as much.
The latent is encoded by quantizing the running residual: starting from $r_0 = z$, at each depth $m = 1,\dots,M$ we select the code that best approximates the current residual vector and subtract it. In other words, at each depth $m$:
\begin{equation}
  k_m = \mathcal{Q}(r_{m-1};\mathcal{C}_m), \qquad
  r_m \;=\; r_{m-1} - e^{m}_{k_m},
  \label{eq:rq-recursion}
\end{equation}
where $\mathcal{C}_m = \{e^{m}_k\}_{k=1}^{K}$ is the codebook at such depth.
The input is thus mapped to the sequence of indices $(k_1,\dots,k_M)$, and the quantized latent is recovered as the cumulative sum $\hat{z} = \sum_{m=1}^{M} e^{m}_{k_m}$.
The reconstructed input $\hat{x} = G(\hat{z})$ is produced by a decoder $G$. 
The tokenizer consists of a self-attention transformer encoder $E$~\citep{vaswaniAttentionAllYou2023}, the residual-quantization module, and a transformer decoder $G$, and is trained end-to-end using the reconstruction objective.
Further details on architecture and training methods are available in the Appendix~\ref{app:rqvae}.

Given the frozen tokenizer, each particle is mapped to its ordered sequence of $M$ residual codes such that an event of $N$ particles is flattened into a sequence of $NM$ tokens, event generation thus becomes a sequence-modeling problem.

\subsection{Autoregressive Generative Model}
\label{sec:armodel}
The core model is a decoder-only transformer with causal self-attention and gated feed-forward networks \citep{shazeer_glu_2020}. A single linear head maps the final hidden state at each position to a categorical distribution over the code vocabulary. 
Further details on architecture specifications and training procedure are reported in Appendix~\ref{app:ar}.
\\We model the distribution over \pflow tokens autoregressively with the transformer, conditioned on the \truth-level event: both \tr and \pf events are tokenized by the same frozen RQ-VAE.  The model's input is the concatenation of the tokenized \tr event and the tokenized \pf event, with each section following its own \texttt{[SOS]} token; the \truth block thus acts as a prefix on which the target block is conditioned. The embedding of each token is the sum of three learned components: a shared embedding over the vocabulary of code indices, a depth embedding indexed by the residual level $m\in\{1,\dots,M\}$, and a learnable linear projection of the corresponding codebook vector $e^{m}_{k}$ itself.

Since events have variable particle multiplicity, the number of reconstructed particles $N_\text{PF}$ is predicted directly from the pooled truth prefix by a feed forward network that is trained to predict logits for the different cardinality amounts. This target multiplicity is supplied to the model through an additional learned ''countdown'' embedding that encodes the number of particles still to be generated at each position. The model is trained on next-token-prediction and cardinality objectives using a composite Cross-Entropy Loss function:
\begin{equation}
    \mathcal{L} =\mathcal{L}^\text{CE}+ \lambda_\text{card}\,\mathcal{L}^\text{CE}_{\text{card}}\,\,\,\,\,, \,\,\,\,\lambda_{\text{card}}=0.1
\end{equation}
After training, to generate an event we tokenize the truth input, prefill it as the prefix, and sample the reconstructed multiplicity $N_\text{PF}$ from the cardinality head's output distribution. We then autoregressively sample exactly $N_\text{PF}\cdot M$ tokens from the categorical distribution. The resulting token sequence is passed through the frozen RQ-VAE decoder to recover the particle-level quantities. Training and inference details are provided in Appendix~\ref{app:training_ar}.

\section{Results}
\label{sec:results}
We evaluate the reconstruction and generation performance on particle, event, and jet-level quantities.
Jets are reconstructed from collimated sprays of particles by clustering nearby constituents in the $\eta-\phi$ plane into composite objects commonly used for downstream analysis in collider experiments.

We use both the held-out in-distribution test set (QCD + $t\bar{t}$) and out-of-distribution datasets ($H \rightarrow 4\ell$ and QCD $1000$) to test the model's ability to generalize beyond physical processes seen during training.
We examine feature distributions, and we also match generated (\fastsim, \fs) and \pflow particles and jets with \truth ones, using the Hungarian matching algorithm~\citep{HUNGARIAN} with $\Delta R = \sqrt{\Delta \eta^2 + \Delta \phi^2}$ as the cost function, and then evaluate the per-feature residual distributions.

The choice of evaluating on the residual sets is intentional: the conditional generation objective is to model the stochastic detector response and reconstruction. 
\fastsim samples should be evaluated in terms of how well they reproduce the full simulation and reconstruction pipeline's response to the incoming detector-stable particles, rather than their ability to deterministically reproduce the exact \pflow features.

\subsection{Tokenizer fidelity and codebook utilization}
\label{sec:tokenizer_perf}

\begin{figure}[h]
\centering
\includegraphics[width = 0.99\linewidth]{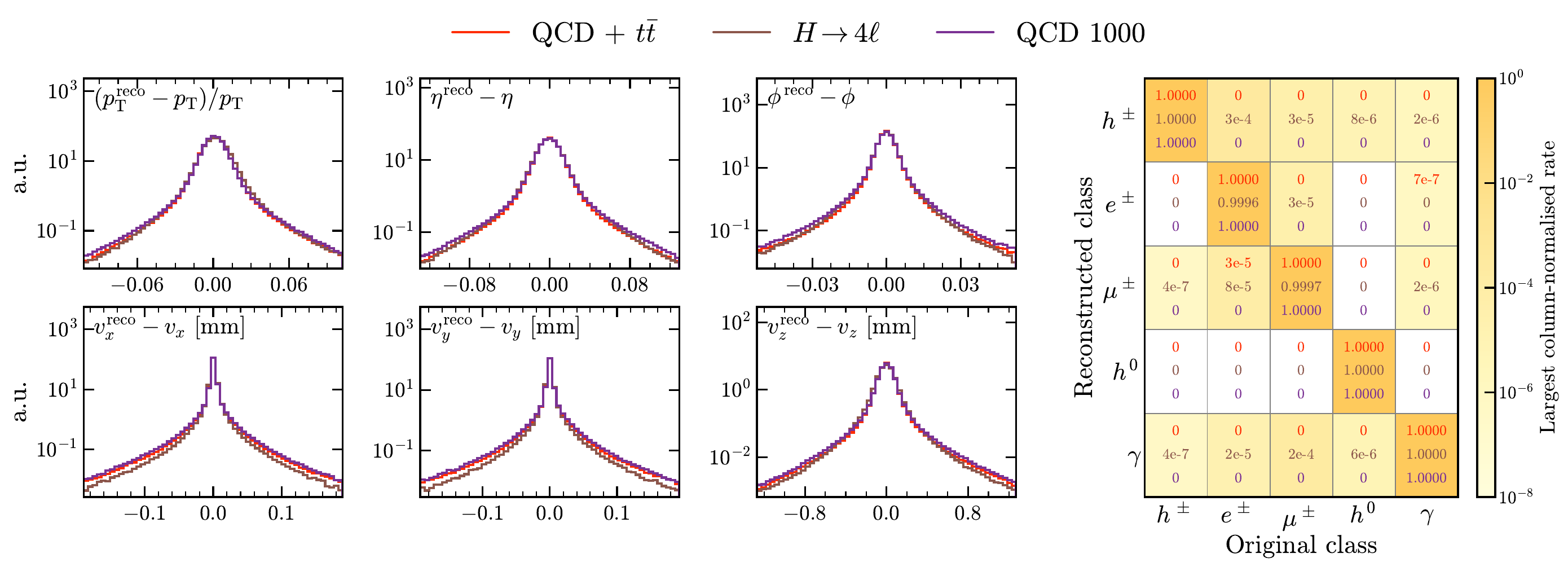}
\caption{Reconstructed features residual distribution, and particle-class confusion matrix. Vertex residuals are only computed for charged particles.}
\label{fig:reco}
\end{figure}

Achieving good performance on the reconstruction objective is a required test of the fidelity of the discrete representation learned by the tokenizer network. The latent is quantized with $M=4$ residual stages (Eq.~\ref{eq:rq-recursion}), and each codebook layer contains $K=8192$ 16-dimensional vectors, more details can be found in Appendix~\ref{app:rqvae}. For each continuous feature, the joint distribution, over both \tr and \pf-level events in the test dataset, of the original and reconstructed values is shown in Figure~\ref{fig:reco}, together with the particle class confusion matrix.

Next we examine how the available discrete representation capacity is utilized across the residual-quantization levels. Figure~\ref{fig:codebook_usage} (Appendix~\ref{app:plots}) shows the code distribution at each depth $m$. For each codebook, we also characterize its usage (i.e. the percentage of codes used at least once) and perplexity; metrics are reported in Table~\ref{tab:code_usage}.

\begin{table}[h]
\caption{Codebook usage and perplexity (PPL) at each residual depth $m$, for the three test datasets.}
\label{tab:code_usage}
\centering
\small
\renewcommand{\arraystretch}{1.15}
\setlength{\tabcolsep}{5pt}
\begin{tabular}{l cc cc cc cc}
 & \multicolumn{2}{c}{$m=1$} & \multicolumn{2}{c}{$m=2$} & \multicolumn{2}{c}{$m=3$} & \multicolumn{2}{c}{$m=4$} \\
\textbf{Dataset} & Usage & Perplexity & Usage & Perplexity & Usage & Perplexity & Usage & Perplexity \\ \hline \\[-6pt]
QCD + $t\bar{t}$ & 58.5\% & 4042 & 99.9\%  & 5575 & 99.9\%  & 4894 & 99.9\%  & 4723 \\
$H \rightarrow 4\ell$ & 58.2\% & 3338 & 99.3\%  & 5382 & 99.4\%  & 4797 & 99.6\%  & 4662 \\
QCD 1000            & 58.5\% & 3897 & 99.9\%  & 5652 & 100.0\% & 4981 & 100.0\% & 4871 \\
\end{tabular}
\end{table}

\subsection{Generation Performance}
\label{sec:gen-perf}
We show the performance of the largest model configuration, trained for four epochs on the full dataset. A performance comparison to other ablations is shown in Table~\ref{tab:w1_models}, in the Appendix~\ref{app:plots}.

\begin{figure}[h]
\centering
\includegraphics[width = \linewidth]{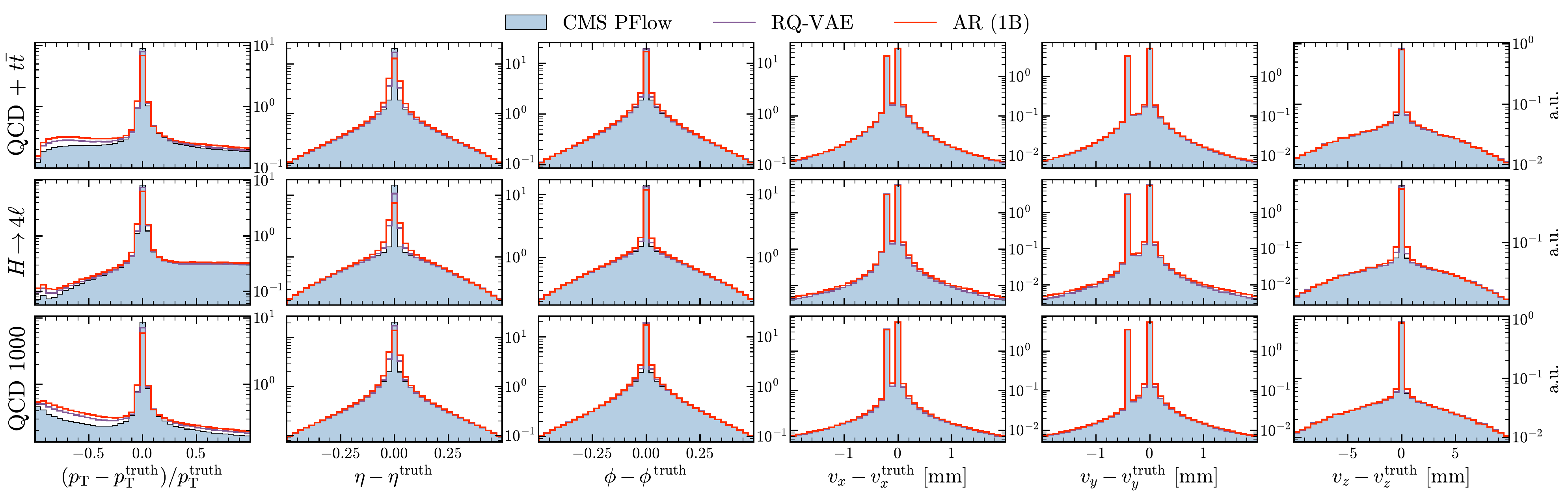}
\caption{Spectra of particle residual features for all three testing datasets. All histograms are normalized to unity; rows correspond to different datasets/processes while columns correspond to different observables.}
\label{fig:particle_feats}
\end{figure}

\paragraph{Particle-level}
Residual distributions of particle-level features, for \pf and \fs particles matched to \tr, are reported in Figure~\ref{fig:particle_feats}. 
The generated residual distributions closely track the corresponding \pf distributions.
The $\eta$ and $\phi$ marginals in Figure~\ref{fig:part_marginals} (Appendix~\ref{app:plots}) also reproduce the detector-dependent structure across the tested datasets.

\paragraph{Event-level}
Event-level variables are a relevant test of the model's ability to capture global event correlations, in particular in an autoregressive setting where the bulk of the generated structure follows from a subset of initially generated hard particles.
For event-level metrics, we consider number of particles,  missing transverse energy in the $x$ and $y$ directions and the scalar sum of particle $p_T$. Residual distributions of these metrics are shown in Figure~\ref{fig:event_feats}, and marginals are displayed in the appendix (Figure~\ref{fig:event_marginals}).

\begin{figure}[h]
\centering
\includegraphics[width = 0.99\linewidth]{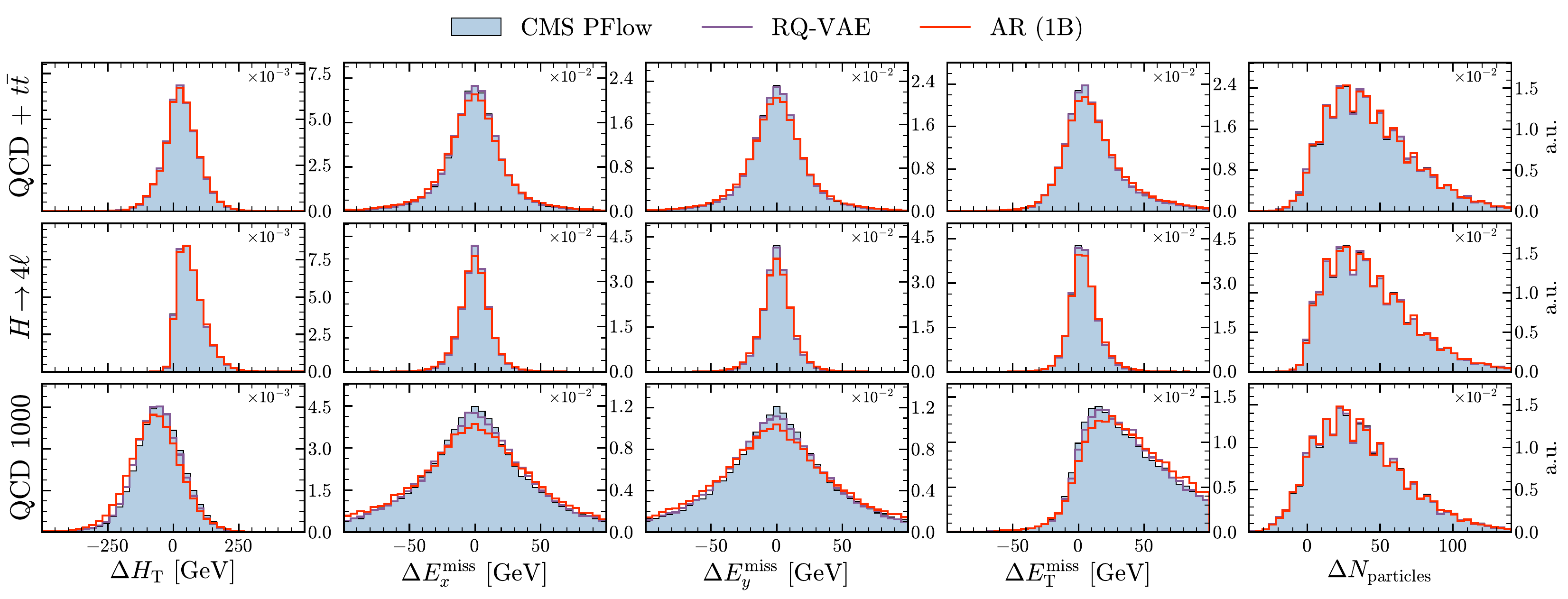} 

\caption{Spectra of event-level residual features for all three testing datasets. All histograms are normalized to unity; rows correspond to different datasets/processes while columns correspond to different observables.}
\label{fig:event_feats}
\end{figure}

\paragraph{Jet-level}
We examine the kinematic properties of the clustered jets ($p_T$, $\eta$, $\phi$) and the jet substructure quantities $C_2$ and $D_2$~\citep{larkoskiEnergyCorrelationFunctions2013, larkoskiPowerCountingBetter2014}. 
While inclusive marginal distributions test the overall kinematic phase-space coverage of the generated \fs samples against target \pf, residual distributions with respect to truth-matched jets evaluate the conditional response and resolution. 
In addition, the substructure observables test the effective modeling of intra-jet particle correlations and detector effects.
Residual distributions for the selected variables are shown in Figure~\ref{fig:jet_feats}, and marginals in the appendix (Figure~\ref{fig:jet_marginals}).

\begin{figure}[h]
\centering

\includegraphics[width = \linewidth]{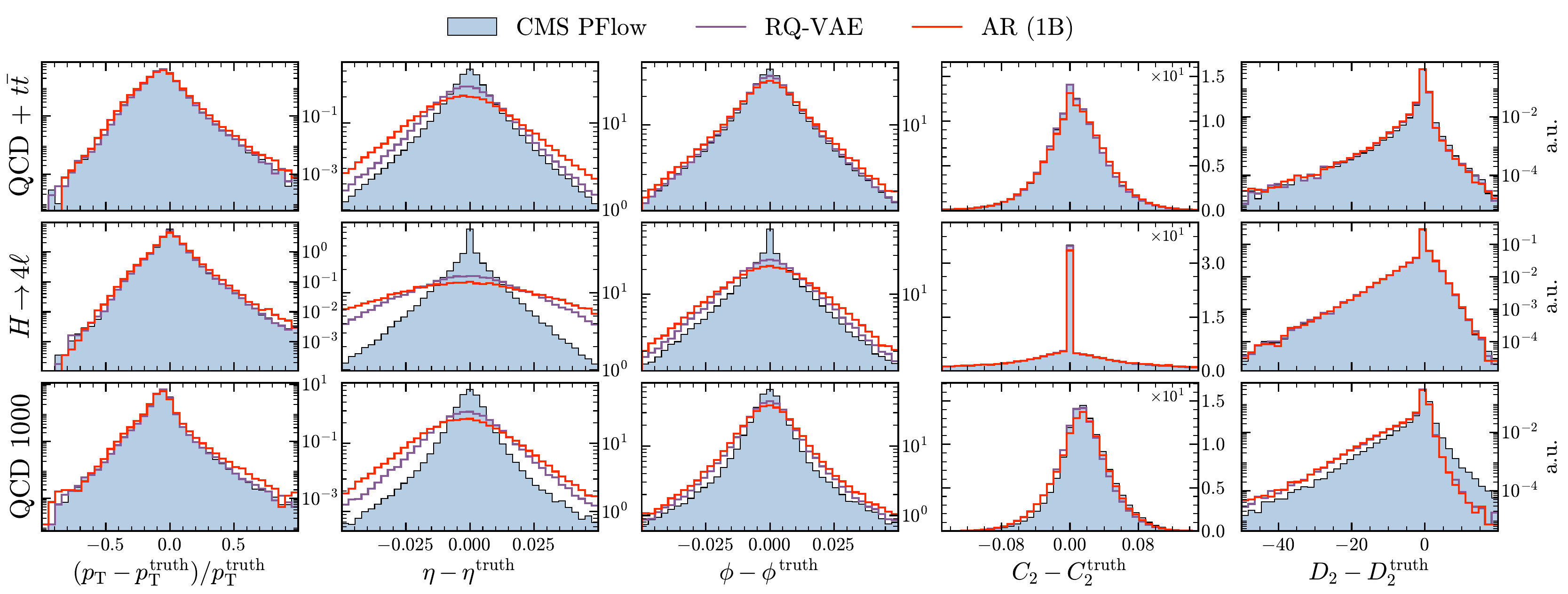}
\caption{Spectra of residuals of jet-level features for all three testing datasets. The residuals are computed against matched truth jets. All histograms are normalized to unity; rows correspond to different datasets/processes while columns correspond to different observables.}
\label{fig:jet_feats}
\end{figure}

\subsection{Scaling study}
\label{sec:scaling}
We study how the generator model's performance changes with model size, training-set size, and repeated data exposure. We evaluate the model on the token cross-entropy loss on the statistically identical set of held-out events and fit its dependence on these training variables.
We then evaluate how scaling affects the generated distributions of physical observables.
\\A full overview of training ablations used for the study is available in Tables~\ref{tab:scaling_models} and~\ref{tab:scaling_data}. 

\begin{table}[ht]
\caption{Model configurations used in the scaling study. Each model is trained with eight data and repetition settings, giving 48 configurations.}
\label{tab:scaling_models}
\begin{center}
\renewcommand{\arraystretch}{1.15}
\setlength{\tabcolsep}{9pt}
\begin{tabular}{lrrrrrr}
\multicolumn{1}{c}{\textbf{Name}} &
\multicolumn{1}{c}{$\boldsymbol{d_{\text{model}}}$} &
\multicolumn{1}{c}{\textbf{Layers}} &
\multicolumn{1}{c}{\textbf{Heads}} &
\multicolumn{1}{c}{$\boldsymbol{d_{\text{head}}}$} &
\multicolumn{1}{c}{$\boldsymbol{d_{\text{ff}}}$} &
\multicolumn{1}{c}{\textbf{Total Params}}
\\ \hline \\[-6pt]
10M  & 256     & 8  & 8  & 32 & 512     & $9.83$M \\
22M  & 384     & 10 & 8  & 48 & 768     & $21.67$M \\
41M  & 512     & 12 & 8  & 64 & 1{,}024 & $40.75$M \\
97M  & 768     & 14 & 12 & 64 & 1{,}536 & $96.71$M \\
187M & 1{,}024 & 16 & 16 & 64 & 2{,}048 & $186.87$M \\
1B   & 2{,}048 & 24 & 32 & 64 & 4{,}096 & $1{,}046.92$M \\
\end{tabular}
\end{center}
\end{table}

\begin{table}[h]
\caption{Data and repetition settings used in the scaling study. Each setting is
applied to all six model sizes. Data fractions are subsampled from the full dataset, in order to preserve the same statistical properties.}
\label{tab:scaling_data}
\begin{center}
\renewcommand{\arraystretch}{1.15}
\setlength{\tabcolsep}{9pt}
\begin{tabular}{rrcrr}
\multicolumn{1}{c}{\textbf{Data fraction}} &
\multicolumn{1}{c}{\textbf{Unique events}} &
\multicolumn{1}{c}{\textbf{Epochs}} &
\multicolumn{1}{c}{\textbf{PF tokens}} &
\multicolumn{1}{c}{\textbf{Total tokens}}
\\ \hline \\[-6pt]
3.3\%   & $546{,}523$      & $1,\ 10,\ 30$ & $324$M  & $749$M \\
33.3\%  & $5{,}514{,}918$  & $1,\ 3,\ 12$ & $3.27$B & $7.56$B \\
100\%   & $16{,}561{,}317$ & $1,\ 4$      & $9.82$B & $22.7$B \\
\end{tabular}
\end{center}
\end{table}

\subsubsection{Loss scaling}
\label{sec:loss_scaling}

The fitted quantity is the held-out next-token-prediction (NTP) cross-entropy $\mathcal{L}$, in nats. This is the token component of the objective defined in Section~\ref{sec:armodel}; the cardinality term is excluded.
First, we ignore data repetition and only consider fresh-data runs in which each event is only shown to the model once at most.
In this regime, we fit the additive Chinchilla~\citep{hoffmann2022trainingcomputeoptimallargelanguage} and coupled Skaling~\citep{videau2026skalingchinchillasexponentsmeet} forms that model the data-constraint free scaling behavior of the model family.

In settings such as detector simulation, generating MC training datasets can be an expensive endeavor. It is important to model the effects of data repetition on ML-based surrogates, and maximize the usefulness of the available synthetic datasets. 
To evaluate the effects of repeated data exposure, we also fit the "Prescriptive" repetition penalty introduced by \citet{lovelace2026prescriptivescalinglawsdata}, and a custom combined form that applies this penalty to the Skaling coupled expression.
We fit the measured losses by unweighted least squares. Table~\ref{tab:scaling_loss_fits} reports the fitted $\alpha, \beta$ parameters and the exponents for compute-optimal allocation $a=\beta/(\alpha+\beta)$ and $b=\alpha/(\alpha+\beta)$ \citep{hoffmann2022trainingcomputeoptimallargelanguage}, together with mean absolute percentage error (MAPE) and coefficient of determination $R^2$, to indicate goodness-of-fit.
Although the fitted parameters vary, all models predict similar values for the $a$ and $b$ optimal-scaling parameters. The full parameters are reported in Table~\ref{tab:scaling_params}, Appendix~\ref{app:loss_scaling}.

\begin{table}[h]
\caption{Fits of the neural scaling laws. Chinchilla and Skaling are fitted to the 18 single-epoch runs, the repetition-aware laws to all 48 runs. Uncertainties are $1\sigma$ errors over refits with one run held-out.}
\label{tab:scaling_loss_fits}
\begin{center}
\small
\renewcommand{\arraystretch}{1.3}
\setlength{\tabcolsep}{3pt}
\resizebox{\linewidth}{!}{%
\begin{tabular}{llccccrr}
\multicolumn{1}{c}{\textbf{Law}} &
\multicolumn{1}{c}{\textbf{Functional form}} &
\multicolumn{1}{c}{\boldmath$\alpha$} &
\multicolumn{1}{c}{\boldmath$\beta$} &
\multicolumn{1}{c}{\boldmath$a$} &
\multicolumn{1}{c}{\boldmath$b$} &
\multicolumn{1}{c}{\textbf{MAPE [\%]}} &
\multicolumn{1}{c}{\boldmath$R^2$}
\\ \hline \\[-6pt]

Chinchilla &
$L_\infty+A N^{-\alpha}+B U^{-\beta}$ &
$\vpm{0.42}{0.18}$ & $\vpm{0.62}{0.15}$ & $\vpm{0.60}{0.14}$ & $\vpm{0.40}{0.14}$ & 3.55 & 0.9695 \\

Skaling &
$L_\infty+\left(A N^{-\alpha}+B U^{-\beta}\right)^k$ &
$\vpm{0.241}{0.052}$ & $\vpm{0.345}{0.070}$ & $\vpm{0.589}{0.048}$ & $\vpm{0.411}{0.048}$ & 1.17 & 0.9961 \\

Prescriptive &
$L_\infty+A N^{-\alpha}+B[U(1+R)]^{-\beta}+P R^\delta\left(\frac{N}{U^\gamma}\right)^\kappa$ &
$\vpm{0.410}{0.092}$ & $\vpm{0.598}{0.036}$ & $\vpm{0.593}{0.074}$ & $\vpm{0.407}{0.074}$ & 2.47 & 0.9690 \\

Custom &
$L_\infty+\left(A N^{-\alpha}+B[U(1+R)]^{-\beta}\right)^k +P R^\delta\left(\frac{N}{U^\gamma}\right)^\kappa$ &
$\vpm{0.273}{0.017}$ & $\vpm{0.349}{0.016}$ & $\vpm{0.561}{0.015}$ & $\vpm{0.439}{0.015}$ & 0.79 & 0.9967 \\

\end{tabular}%
}
\end{center}
\end{table}

\begin{figure}[h]
\centering

\includegraphics[width = \linewidth]{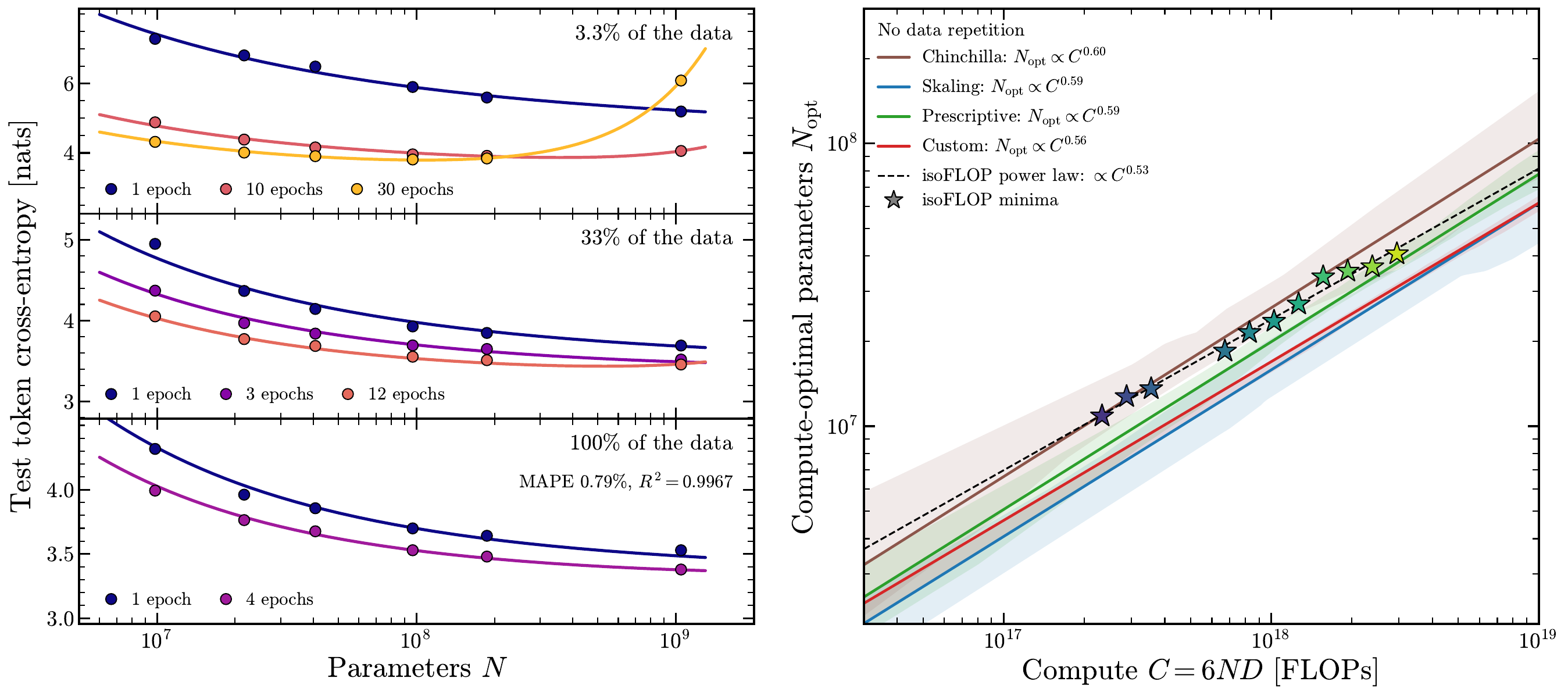}

\caption{Left: each trained model's NTP loss against its parameter count. The lines show the custom fitted law's prediction for the given combination of dataset size and repetition. Right: compute-guided fit-predicted optimal scaling behaviors and isoFLOP curves minima. The methodology for recovering these minima is discussed in the Appendix~\ref{app:isoflop}.}
\label{fig:loss_fit_and_isoFLOP}
\end{figure}

Figure~\ref{fig:loss_fit_and_isoFLOP} compares the measured held-out losses with the fitted curves and the implied compute-optimal scaling of $N$.
Larger models overfit sooner on smaller datasets but achieve lower loss when sufficient unique data are available.
Over the compute range covered by the single-epoch runs, the fitted compute-optimal model size remains below the largest models in our grid, indicating that the study is data-constrained.
The isoFLOP-fitted analysis yields $N\propto C^{0.53}$.

\subsubsection{Scaling of physical metrics}
\label{sec:physical_scaling}
It is not immediately obvious how modeling of the decoded physical variables correlates with a lower value of the NTP Loss. The model might produce token sequences that appear satisfactory on a surface level but produce nonsensical physical results once decoded. For this reason it is important to choose metrics that test both local and global features of the generated samples.

\begin{figure}[h]
\centering
\includegraphics[width = \linewidth]{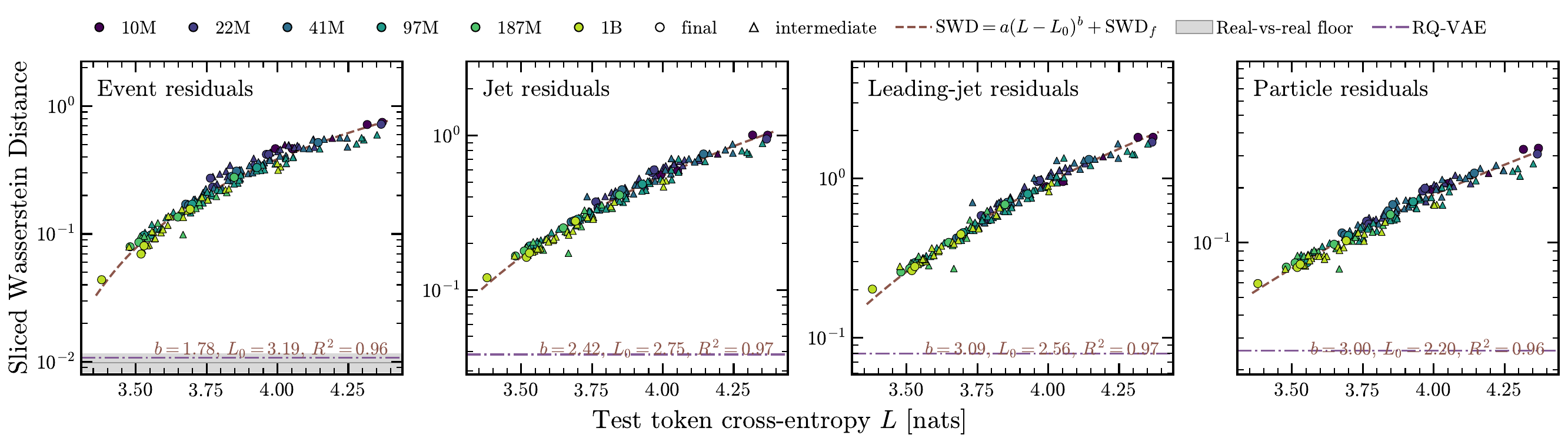}

\caption{Measured Sliced Wasserstein Distance for all valid (selection criteria in Appendix~\ref{app:physical_scaling}) runs, and fitted law in the form of Equation~\ref{eq:metric_law}. We include both final and intermediate checkpoints and show the fitted power law parameters for each feature.}
\label{fig:SWD_plots}
\end{figure}

\paragraph{Global point-cloud metrics}
First, we adopt a procedure similar to that of \citet{amram2026neuralscalinglawsjet} by comparing the NTP loss with the Sliced Wasserstein Distance (SWD) \citep{julien:hal-00476064, bonneel:hal-00881872} between \fs and \pf samples, but we instead evaluate on the particle-, event-, and jet-level \textbf{residual} feature manifolds.
For each feature group, we subsample the point clouds and average the $W_1$ Wasserstein Distance over $256$ random projections after standardising every feature with the mean and standard deviation of the \pf reference. We also include the statistical noise barrier obtained by computing the SWD between statistically identical samples. 
\\In Figure~\ref{fig:SWD_plots} we plot the NTP loss for given checkpoints against the SWD on the given feature sets of the associated produced sample from the testing datasets. We find, by fitting a power law of the type:
\begin{equation}\label{eq:metric_law}
    \text{SWD} = a(\mathcal{L}-\mathcal{L}_0)^b + \text{SWD}_f
\end{equation}
that the loss achieved on the held-out test dataset is a very strong performance predictor on this global metric. Fit results and a visualization are available in Figure~\ref{fig:SWD_plots}. Details on the full procedure are available in Appendix~\ref{app:physical_scaling}.

\begin{figure}[h]
\centering
\includegraphics[width = \linewidth]{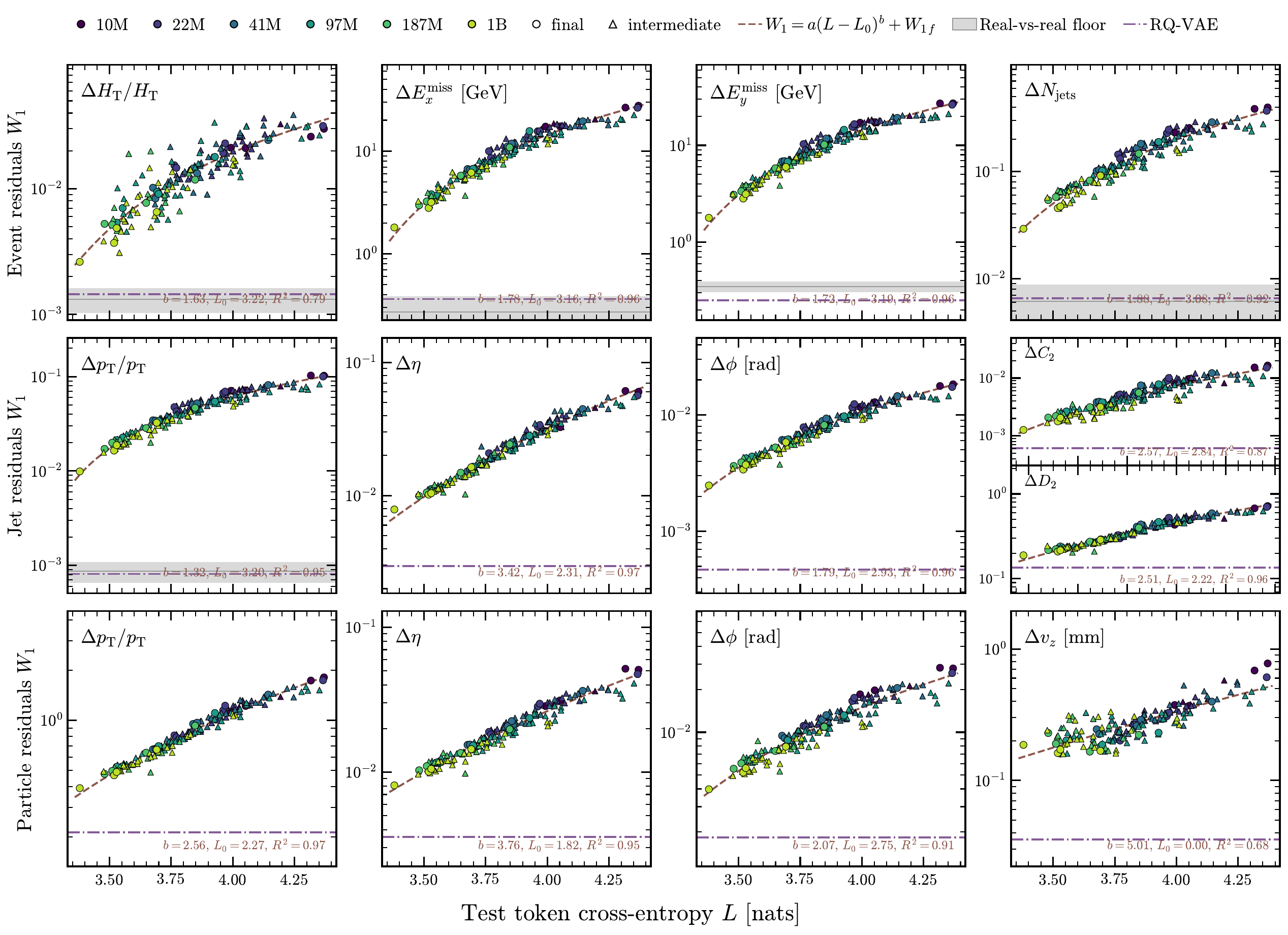}

\caption{Per-feature $W_1$ Wasserstein Distance between the \pf and \fs valid (selection criteria in Appendix~\ref{app:physical_scaling}) residual distributions. We include both final and intermediate checkpoints and show the fitted power law parameters for each feature.}
\label{fig:WD_res_plots}
\end{figure}

\paragraph{Individual metrics}
Observing scaling behavior on distributions of the individual output features is not a trivial task: modeling the marginal distributions of these is often easier and can result in differences that fall within two-sample noise. We find, however, that each one of the individual residual feature distributions exhibits scaling behavior, in the same form as Eq.~\ref{eq:metric_law}, as the global feature manifolds, and that the achieved loss value is also a strong individual predictor of the fidelity of these generated physical metrics. 
In Figure~\ref{fig:WD_res_plots} we check the one-dimensional $W_1$ Wasserstein Distance between the \fastsim and \pflow residual distributions, and show the fit results across the different features. We find that improvements in achieved loss transfer to the individual features.

\section{Conclusion}
We apply discrete autoregressive modeling to conditional full-event simulation, demonstrating how residual-quantized representations of point-clouds enable generation of particle-level detector responses. By employing a frozen RQ-VAE tokenizer with an autoregressive transformer, we recover continuous particle-level outputs from discrete sequences while preserving the flexibility of full-event generation. We demonstrate the fidelity of the learned representation and evaluate the generated events across particle, event, and jet-level observables, including on physical processes not encountered during training.

We then study the scaling behavior of this model family across a range of model and dataset sizes, find that the next-token prediction loss exhibits familiar scaling characteristics, and that the effects of data repetition can be captured by a repetition-aware scaling law. 

Our empirical results demonstrate a direct link between token-level modeling performance and the fidelity of generated physical observables.
We show that the achieved next-token prediction loss is a strong predictor of both global distributional distances and individual metrics, with their relationship well described by a simple power law. These findings suggest that token-level scaling laws may provide a useful proxy for estimating downstream physical performance, validate residual-quantized autoregressive architectures for scalable simulation in high-energy physics and demonstrate how empirical scaling laws can guide compute allocation for ML surrogate models.

\subsection*{Acknowledgments}
This work has been supported by the Minerva Grant, the Knell Family Institute for Artificial Intelligence, The Benoziyo Center for High Energy Physics, a BSF Grant, and the Krenter-Perinot Center for High Energy Particle Physics.

\newpage
\bibliographystyle{unsrtnat}
\bibliography{scaling}

@misc{h4lep_data,
  doi = {10.7483/OPENDATA.CMS.G4SB.Y2BX},
  url = {http://opendata.cern.ch/record/1443},
  author = {{CMS Collaboration}},
  title = {{Simulated dataset GluGluToHToZZTo4L\_M-125\_7TeV-minloHJJ-pythia6-tauola in AODSIM format for 2011 collision data (SM Higgs)}},
  publisher = {CERN Open Data Portal},
  year = {2016}
}

@misc{ttbar_data,
  doi = {10.7483/OPENDATA.CMS.VNS2.C226},
  url = {http://opendata.cern.ch/record/1662},
  author = {{CMS Collaboration}},
  title = {{Simulated dataset TTJets\_MSDecays\_central\_TuneZ2\_7TeV-madgraph-tauola in AODSIM format for 2011 collision data (SM Inclusive)}},
  publisher = {CERN Open Data Portal},
  year = {2016}
}

@misc{qcd470_data,
  doi = {10.7483/OPENDATA.CMS.BKTD.SGJX},
  url = {http://opendata.cern.ch/record/1553},
  author = {{CMS Collaboration}},
  title = {{Simulated dataset QCD\_Pt-470to600\_TuneZ2\_7TeV\_pythia6 in AODSIM format for 2011 collision data (SM Exclusive)}},
  publisher = {CERN Open Data Portal},
  year = {2016}
}

@misc{qcd1000_data,
  doi = {10.7483/OPENDATA.CMS.96U2.3YAH},
  url = {http://opendata.cern.ch/record/1342},
  author = {{CMS Collaboration}},
  title = {{Simulated dataset QCD\_Pt-1000to1400\_TuneZ2\_7TeV\_pythia6 in AODSIM format for 2011 collision data (SM Exclusive)}},
  publisher = {CERN Open Data Portal},
  year = {2016}
}

@misc{qcd600_data,
  doi = {10.7483/OPENDATA.CMS.EJT7.KSAY},
  url = {http://opendata.cern.ch/record/1558},
  author = {{CMS Collaboration}},
  title = {{Simulated dataset QCD\_Pt-600to800\_TuneZ2\_7TeV\_pythia6 in AODSIM format for 2011 collision data (SM Exclusive)}},
  publisher = {CERN Open Data Portal},
  year = {2016}
}

@article{agostinelliGeant4aSimulationToolkit2003,
    title = {Geant4—a simulation toolkit},
    volume = {506},
    issn = {0168-9002},
    url = {https://www.sciencedirect.com/science/article/pii/S0168900203013688},
    doi = {10.1016/S0168-9002(03)01368-8},
    number = {3},
    urldate = {2024-04-15},
    journal = {Nuclear Instruments and Methods in Physics Research Section A: Accelerators, Spectrometers, Detectors and Associated Equipment},
    author = {Agostinelli, S. and Allison, J. and Amako, K. and Apostolakis, J. and Araujo, H. and Arce, P. and Asai, M. and Axen, D. and Banerjee, S. and Barrand, G. and Behner, F. and Bellagamba, L. and Boudreau, J. and Broglia, L. and Brunengo, A. and Burkhardt, H. and Chauvie, S. and Chuma, J. and Chytracek, R. and Cooperman, G. and Cosmo, G. and Degtyarenko, P. and Dell'Acqua, A. and Depaola, G. and Dietrich, D. and Enami, R. and Feliciello, A. and Ferguson, C. and Fesefeldt, H. and Folger, G. and Foppiano, F. and Forti, A. and Garelli, S. and Giani, S. and Giannitrapani, R. and Gibin, D. and Gómez Cadenas, J. J. and González, I. and Gracia Abril, G. and Greeniaus, G. and Greiner, W. and Grichine, V. and Grossheim, A. and Guatelli, S. and Gumplinger, P. and Hamatsu, R. and Hashimoto, K. and Hasui, H. and Heikkinen, A. and Howard, A. and Ivanchenko, V. and Johnson, A. and Jones, F. W. and Kallenbach, J. and Kanaya, N. and Kawabata, M. and Kawabata, Y. and Kawaguti, M. and Kelner, S. and Kent, P. and Kimura, A. and Kodama, T. and Kokoulin, R. and Kossov, M. and Kurashige, H. and Lamanna, E. and Lampén, T. and Lara, V. and Lefebure, V. and Lei, F. and Liendl, M. and Lockman, W. and Longo, F. and Magni, S. and Maire, M. and Medernach, E. and Minamimoto, K. and Mora de Freitas, P. and Morita, Y. and Murakami, K. and Nagamatu, M. and Nartallo, R. and Nieminen, P. and Nishimura, T. and Ohtsubo, K. and Okamura, M. and O'Neale, S. and Oohata, Y. and Paech, K. and Perl, J. and Pfeiffer, A. and Pia, M. G. and Ranjard, F. and Rybin, A. and Sadilov, S. and Di Salvo, E. and Santin, G. and Sasaki, T. and Savvas, N. and Sawada, Y. and Scherer, S. and Sei, S. and Sirotenko, V. and Smith, D. and Starkov, N. and Stoecker, H. and Sulkimo, J. and Takahata, M. and Tanaka, S. and Tcherniaev, E. and Safai Tehrani, E. and Tropeano, M. and Truscott, P. and Uno, H. and Urban, L. and Urban, P. and Verderi, M. and Walkden, A. and Wander, W. and Weber, H. and Wellisch, J. P. and Wenaus, T. and Williams, D. C. and Wright, D. and Yamada, T. and Yoshida, H. and Zschiesche, D.},
    month = jul,
    year = {2003},
    pages = {250--303},
}

@inproceedings{flashattn,
author = {Dao, Tri and Fu, Daniel Y. and Ermon, Stefano and Rudra, Atri and R{\'e}, Christopher},
title = {FLASHATTENTION: fast and memory-efficient exact attention with IO-awareness},
year = {2022},
isbn = {9781713871088},
publisher = {Curran Associates Inc.},
address = {Red Hook, NY, USA},
booktitle = {Proceedings of the 36th International Conference on Neural Information Processing Systems},
articleno = {1189},
numpages = {16},
location = {New Orleans, LA, USA},
series = {NIPS '22}
}

@misc{dao2023flashattention2fasterattentionbetter,
      title={FlashAttention-2: Faster Attention with Better Parallelism and Work Partitioning}, 
      author={Tri Dao},
      year={2023},
      eprint={2307.08691},
      archivePrefix={arXiv},
      primaryClass={cs.LG},
      url={https://arxiv.org/abs/2307.08691}, 
}

@misc{loshchilov_decoupled_2019,
    title = {Decoupled {Weight} {Decay} {Regularization}},
    url = {http://arxiv.org/abs/1711.05101},
    doi = {10.48550/arXiv.1711.05101},
    urldate = {2026-09-24},
    publisher = {arXiv},
    author = {Loshchilov, Ilya and Hutter, Frank},
    month = jan,
    year = {2019},
    note = {arXiv:1711.05101 [cs.LG]},
}

@misc{zhuoHybridNormStableEfficient2025,
    title = {{HybridNorm}: {Towards} {Stable} and {Efficient} {Transformer} {Training} via {Hybrid} {Normalization}},
    shorttitle = {{HybridNorm}},
    url = {http://arxiv.org/abs/2503.04598},
    doi = {10.48550/arXiv.2503.04598},
    urldate = {2025-03-07},
    publisher = {arXiv},
    author = {Zhuo, Zhijian and Zeng, Yutao and Wang, Ya and Zhang, Sijun and Yang, Jian and Li, Xiaoqing and Zhou, Xun and Ma, Jinwen},
    month = mar,
    year = {2025},
    note = {arXiv:2503.04598 [cs]},
}

@misc{su_roformer_2023,
    title = {{RoFormer}: {Enhanced} {Transformer} with {Rotary} {Position} {Embedding}},
    shorttitle = {{RoFormer}},
    url = {http://arxiv.org/abs/2104.09864},
    doi = {10.48550/arXiv.2104.09864},
    urldate = {2026-09-24},
    publisher = {arXiv},
    author = {Su, Jianlin and Lu, Yu and Pan, Shengfeng and Murtadha, Ahmed and Wen, Bo and Liu, Yunfeng},
    month = nov,
    year = {2023},
    note = {arXiv:2104.09864 [cs.CL]},
}

@misc{shazeer_glu_2020,
    title = {{GLU} {Variants} {Improve} {Transformer}},
    url = {http://arxiv.org/abs/2002.05202},
    doi = {10.48550/arXiv.2002.05202},
    urldate = {2026-09-24},
    publisher = {arXiv},
    author = {Shazeer, Noam},
    month = feb,
    year = {2020},
    note = {arXiv:2002.05202 [cs.LG]},
}

@misc{dehghaniScalingVisionTransformers2023,
    title = {Scaling {Vision} {Transformers} to 22 {Billion} {Parameters}},
    url = {http://arxiv.org/abs/2302.05442},
    doi = {10.48550/arXiv.2302.05442},
    urldate = {2025-08-22},
    publisher = {arXiv},
    author = {Dehghani, Mostafa and Djolonga, Josip and Mustafa, Basil and Padlewski, Piotr and Heek, Jonathan and Gilmer, Justin and Steiner, Andreas and Caron, Mathilde and Geirhos, Robert and Alabdulmohsin, Ibrahim and Jenatton, Rodolphe and Beyer, Lucas and Tschannen, Michael and Arnab, Anurag and Wang, Xiao and Riquelme, Carlos and Minderer, Matthias and Puigcerver, Joan and Evci, Utku and Kumar, Manoj and Steenkiste, Sjoerd van and Elsayed, Gamaleldin F. and Mahendran, Aravindh and Yu, Fisher and Oliver, Avital and Huot, Fantine and Bastings, Jasmijn and Collier, Mark Patrick and Gritsenko, Alexey and Birodkar, Vighnesh and Vasconcelos, Cristina and Tay, Yi and Mensink, Thomas and Kolesnikov, Alexander and Pavetić, Filip and Tran, Dustin and Kipf, Thomas and Lučić, Mario and Zhai, Xiaohua and Keysers, Daniel and Harmsen, Jeremiah and Houlsby, Neil},
    month = feb,
    year = {2023},
    note = {arXiv:2302.05442 [cs]},
}

@misc{ba_layer_2016,
    title = {Layer {Normalization}},
    url = {http://arxiv.org/abs/1607.06450},
    doi = {10.48550/arXiv.1607.06450},
    urldate = {2026-09-24},
    publisher = {arXiv},
    author = {Ba, Jimmy Lei and Kiros, Jamie Ryan and Hinton, Geoffrey E.},
    month = jul,
    year = {2016},
    note = {arXiv:1607.06450 [stat.ML]},
}

@misc{fifty_restructuring_2024,
    title = {Restructuring {Vector} {Quantization} with the {Rotation} {Trick}},
    url = {https://arxiv.org/abs/2410.06424v2},
    language = {en},
    urldate = {2026-09-24},
    journal = {arXiv.org},
    author = {Fifty, Christopher and Junkins, Ronald G. and Duan, Dennis and Iyengar, Aniketh and Liu, Jerry W. and Amid, Ehsan and Thrun, Sebastian and Ré, Christopher},
    month = oct,
    year = {2024},
}

@misc{zeghidourSoundStreamEndtoEndNeural2021,
    title = {{SoundStream}: {An} {End}-to-{End} {Neural} {Audio} {Codec}},
    shorttitle = {{SoundStream}},
    url = {http://arxiv.org/abs/2107.03312},
    doi = {10.48550/arXiv.2107.03312},
    urldate = {2025-02-01},
    publisher = {arXiv},
    author = {Zeghidour, Neil and Luebs, Alejandro and Omran, Ahmed and Skoglund, Jan and Tagliasacchi, Marco},
    month = jul,
    year = {2021},
    note = {arXiv:2107.03312 [cs]},
}

@misc{oord_neural_2018,
    title = {Neural {Discrete} {Representation} {Learning}},
    url = {http://arxiv.org/abs/1711.00937},
    doi = {10.48550/arXiv.1711.00937},
    urldate = {2026-09-24},
    publisher = {arXiv},
    author = {Oord, Aaron van den and Vinyals, Oriol and Kavukcuoglu, Koray},
    month = may,
    year = {2018},
    note = {arXiv:1711.00937 [cs.LG]},
}

@misc{razavi_generating_2019,
    title = {Generating {Diverse} {High}-{Fidelity} {Images} with {VQ}-{VAE}-2},
    url = {http://arxiv.org/abs/1906.00446},
    doi = {10.48550/arXiv.1906.00446},
    urldate = {2026-09-24},
    publisher = {arXiv},
    author = {Razavi, Ali and Oord, Aaron van den and Vinyals, Oriol},
    month = jun,
    year = {2019},
    note = {arXiv:1906.00446 [cs.LG]},
}

@article{sjostrandBriefIntroductionPYTHIA2008,
    title = {A brief introduction to {PYTHIA} 8.1},
    volume = {178},
    issn = {0010-4655},
    url = {https://www.sciencedirect.com/science/article/pii/S0010465508000441},
    doi = {10.1016/j.cpc.2008.01.036},
    number = {11},
    urldate = {2024-04-14},
    journal = {Computer Physics Communications},
    author = {Sjöstrand, Torbjörn and Mrenna, Stephen and Skands, Peter},
    month = jun,
    year = {2008},
    pages = {852--867},
}

@techreport{CMS-PAS-PFT-09-001,
      author = {{CMS Collaboration}},
      title         = "{Particle-Flow Event Reconstruction in CMS and Performance
                       for Jets, Taus, and MET}",
      institution   = "CERN",
      reportNumber  = "CMS-PAS-PFT-09-001",
      address       = "Geneva",
      year          = "2009",
      url           = "https://cds.cern.ch/record/1194487",
}

@article{dreyerConditionalDeepGenerative2026,
    title = {Conditional deep generative models for simultaneous simulation and reconstruction of entire events},
    volume = {113},
    url = {https://link.aps.org/doi/10.1103/14ph-482n},
    doi = {10.1103/14ph-482n},
    number = {3},
    urldate = {2026-09-24},
    journal = {Physical Review D},
    publisher = {American Physical Society},
    author = {Dreyer, Etienne and Gross, Eilam and Kobylianskii, Dmitrii and Mikuni, Vinicius and Nachman, Benjamin},
    month = feb,
    year = {2026},
    pages = {032005},
}

@article{larkoskiEnergyCorrelationFunctions2013,
    title = {Energy correlation functions for jet substructure},
    volume = {2013},
    issn = {1029-8479},
    url = {https://link.springer.com/10.1007/JHEP06(2013)108},
    doi = {10.1007/JHEP06(2013)108},
    language = {en},
    number = {6},
    urldate = {2024-11-10},
    journal = {Journal of High Energy Physics},
    author = {Larkoski, Andrew J. and Salam, Gavin P. and Thaler, Jesse},
    month = jun,
    year = {2013},
    note = {arXiv:1305.0007},
    pages = {108},
}

@article{butterMachineLearningLHC2023a,
    title = {Machine {Learning} and {LHC} {Event} {Generation}},
    volume = {14},
    issn = {2542-4653},
    url = {http://arxiv.org/abs/2203.07460},
    doi = {10.21468/SciPostPhys.14.4.079},
    number = {4},
    urldate = {2026-09-25},
    journal = {SciPost Physics},
    author = {Butter, Anja and Plehn, Tilman and Schumann, Steffen and Badger, Simon and Caron, Sascha and Cranmer, Kyle and Bello, Francesco Armando Di and Dreyer, Etienne and Forte, Stefano and Ganguly, Sanmay and Gonçalves, Dorival and Gross, Eilam and Heimel, Theo and Heinrich, Gudrun and Heinrich, Lukas and Held, Alexander and Höche, Stefan and Howard, Jessica N. and Ilten, Philip and Isaacson, Joshua and Janßen, Timo and Jones, Stephen and Kado, Marumi and Kagan, Michael and Kasieczka, Gregor and Kling, Felix and Kraml, Sabine and Krause, Claudius and Krauss, Frank and Kröninger, Kevin and Barman, Rahool Kumar and Luchmann, Michel and Magerya, Vitaly and Maitre, Daniel and Malaescu, Bogdan and Maltoni, Fabio and Martini, Till and Mattelaer, Olivier and Nachman, Benjamin and Pitz, Sebastian and Rojo, Juan and Schwartz, Matthew and Shih, David and Siegert, Frank and Stegeman, Roy and Stienen, Bob and Thaler, Jesse and Verheyen, Rob and Whiteson, Daniel and Winterhalder, Ramon and Zupan, Jure},
    month = apr,
    year = {2023},
    note = {arXiv:2203.07460 [hep-ph]},
    pages = {079},
}

@article{krauseCaloChallenge2022Community2025,
    title = {{CaloChallenge} 2022: {A} {Community} {Challenge} for {Fast} {Calorimeter} {Simulation}},
    volume = {88},
    issn = {0034-4885, 1361-6633},
    shorttitle = {{CaloChallenge} 2022},
    url = {http://arxiv.org/abs/2410.21611},
    doi = {10.1088/1361-6633/ae1304},
    number = {11},
    urldate = {2026-09-25},
    journal = {Reports on Progress in Physics},
    author = {Krause, Claudius and Giannelli, Michele Faucci and Kasieczka, Gregor and Nachman, Benjamin and Salamani, Dalila and Shih, David and Zaborowska, Anna and Amram, Oz and Borras, Kerstin and Buckley, Matthew R. and Buhmann, Erik and Buss, Thorsten and Cardoso, Renato Paulo Da Costa and Caterini, Anthony L. and Chernyavskaya, Nadezda and Corchia, Federico A. G. and Cresswell, Jesse C. and Diefenbacher, Sascha and Dreyer, Etienne and Ekambaram, Vijay and Eren, Engin and Ernst, Florian and Favaro, Luigi and Franchini, Matteo and Gaede, Frank and Gross, Eilam and Hsu, Shih-Chieh and Jaruskova, Kristina and Käch, Benno and Kalagnanam, Jayant and Kansal, Raghav and Kim, Taewoo and Kobylianskii, Dmitrii and Korol, Anatolii and Korcari, William and Krücker, Dirk and Krüger, Katja and Letizia, Marco and Li, Shu and Liu, Qibin and Liu, Xiulong and Loaiza-Ganem, Gabriel and Madula, Thandikire and McKeown, Peter and Melzer-Pellmann, Isabell-A. and Mikuni, Vinicius and Nguyen, Nam and Ore, Ayodele and Schweitzer, Sofia Palacios and Pang, Ian and Pedro, Kevin and Plehn, Tilman and Pokorski, Witold and Qu, Huilin and Raikwar, Piyush and Raine, John A. and Reyes-Gonzalez, Humberto and Rinaldi, Lorenzo and Ross, Brendan Leigh and Scham, Moritz A. W. and Schnake, Simon and Shimmin, Chase and Shlizerman, Eli and Soybelman, Nathalie and Srivatsa, Mudhakar and Tsolaki, Kalliopi and Vallecorsa, Sofia and Yeo, Kyongmin and Zhang, Rui},
    month = nov,
    year = {2025},
    note = {arXiv:2410.21611 [physics.ins-det]},
    pages = {116201},
}

@article{aadTransformingJetFlavour2026,
    title = {Transforming jet flavour tagging at {ATLAS}},
    volume = {17},
    copyright = {2025 The Author(s)},
    issn = {2041-1723},
    url = {https://www.nature.com/articles/s41467-025-65059-6},
    doi = {10.1038/s41467-025-65059-6},
    language = {en},
    number = {1},
    urldate = {2026-09-24},
    journal = {Nature Communications},
    author = {{ATLAS Collaboration}},
    month = jan,
    year = {2026},
    pages = {541},
}

@techreport{CERN-LHCC-2022-005,
      collaboration = "ATLAS",
      title         = "{ATLAS Software and Computing HL-LHC Roadmap}",
      institution   = "CERN",
      reportNumber  = "CERN-LHCC-2022-005, LHCC-G-182",
      address       = "Geneva",
      year          = "2022",
      url           = "https://cds.cern.ch/record/2802918",
      author        = "{{ATLAS Collaboration}}"
}

@techreport{Collaboration:2957472,
      author        = "{{CMS Collaboration}}",
      title         = "{CMS Offline Software and Computing for HL-LHC Conceptual
                       Design Report}",
      institution   = "CERN",
      reportNumber  = "CERN-LHCC-2026-003, LHCC-G-186",
      address       = "Geneva",
      year          = "2026",
      url           = "https://cds.cern.ch/record/2957472",
      doi           = "10.17181/CERN.90J3.ZSH1",
}

@article{birkOmniJet$a$FirstCrosstask2024,
    title = {{OmniJet}-$\alpha$: {The} first cross-task foundation model for particle physics},
    volume = {5},
    issn = {2632-2153},
    shorttitle = {{OmniJet}-$\alpha$},
    url = {http://arxiv.org/abs/2403.05618},
    doi = {10.1088/2632-2153/ad66ad},
    number = {3},
    urldate = {2024-12-31},
    journal = {Machine Learning: Science and Technology},
    author = {Birk, Joschka and Hallin, Anna and Kasieczka, Gregor},
    month = sep,
    year = {2024},
    note = {arXiv:2403.05618 [hep-ph]},
    pages = {035031},
}

@article{gollingMaskedParticleModeling2024b,
    title = {Masked {Particle} {Modeling} on {Sets}: {Towards} {Self}-{Supervised} {High} {Energy} {Physics} {Foundation} {Models}},
    volume = {5},
    issn = {2632-2153},
    shorttitle = {Masked {Particle} {Modeling} on {Sets}},
    url = {http://arxiv.org/abs/2401.13537},
    doi = {10.1088/2632-2153/ad64a8},
    number = {3},
    urldate = {2026-09-25},
    journal = {Machine Learning: Science and Technology},
    author = {Golling, Tobias and Heinrich, Lukas and Kagan, Michael and Klein, Samuel and Leigh, Matthew and Osadchy, Margarita and Raine, John Andrew},
    month = sep,
    year = {2024},
    note = {arXiv:2401.13537 [hep-ph]},
    pages = {035074},
}

@misc{birkOmniJet$a_C$LearningPoint2025,
    title = {{OmniJet}-$\alpha\_C$: {Learning} point cloud calorimeter simulations using generative transformers},
    shorttitle = {{OmniJet}-$\alpha\_C$},
    url = {http://arxiv.org/abs/2501.05534},
    doi = {10.48550/arXiv.2501.05534},
    urldate = {2026-09-25},
    publisher = {arXiv},
    author = {Birk, Joschka and Gaede, Frank and Hallin, Anna and Kasieczka, Gregor and Mozzanica, Martina and Rose, Henning},
    month = jun,
    year = {2025},
    note = {arXiv:2501.05534 [hep-ph]},
}

@misc{amramAspenOpenJets2024,
    title = {Aspen {Open} {Jets}: {Unlocking} {LHC} {Data} for {Foundation} {Models} in {Particle} {Physics}},
    shorttitle = {Aspen {Open} {Jets}},
    url = {http://arxiv.org/abs/2412.10504},
    doi = {10.48550/arXiv.2412.10504},
    urldate = {2025-01-07},
    publisher = {arXiv},
    author = {Amram, Oz and Anzalone, Luca and Birk, Joschka and Faroughy, Darius A. and Hallin, Anna and Kasieczka, Gregor and Krämer, Michael and Pang, Ian and Reyes-Gonzalez, Humberto and Shih, David},
    month = dec,
    year = {2024},
    note = {arXiv:2412.10504 [hep-ph]},
}

@article{bhimjiOmniLearnedFoundationModel2026,
    title = {{OmniLearned}: {A} {Foundation} {Model} {Framework} for {All} {Tasks} {Involving} {Jet} {Physics}},
    volume = {113},
    issn = {2470-0010, 2470-0029},
    shorttitle = {{OmniLearned}},
    url = {http://arxiv.org/abs/2510.24066},
    doi = {10.1103/knmd-f5jm},
    number = {3},
    urldate = {2026-09-25},
    journal = {Physical Review D},
    author = {Bhimji, Wahid and Harris, Chris and Mikuni, Vinicius and Nachman, Benjamin},
    month = feb,
    year = {2026},
    note = {arXiv:2510.24066 [hep-ph]},
    pages = {032020},
}

@misc{cardona-giraldoGeneralizableFoundationModels2026,
    title = {Generalizable {Foundation} {Models} for {Calorimetry} via {Mixtures}-of-{Experts} and {Parameter} {Efficient} {Fine} {Tuning}},
    url = {http://arxiv.org/abs/2603.28804},
    doi = {10.48550/arXiv.2603.28804},
    urldate = {2026-09-25},
    publisher = {arXiv},
    author = {Cardona-Giraldo, Carlos and Fanelli, Cristiano and Giroux, James and Granger, Cole and Nachman, Benjamin and Sabin, Gerald},
    month = mar,
    year = {2026},
    note = {arXiv:2603.28804 [physics.ins-det]},
}

@misc{vigl2026neuralscalinglawsboosted,
      title={Neural Scaling Laws for Boosted Jet Tagging}, 
      author={Matthias Vigl and Nicole Hartman and Michael Kagan and Lukas Heinrich},
      year={2026},
      eprint={2602.15781},
      archivePrefix={arXiv},
      primaryClass={hep-ex},
      url={https://arxiv.org/abs/2602.15781}, 
}

@misc{amram2026neuralscalinglawsjet,
      title={Neural Scaling Laws for Jet Generation}, 
      author={Oz Amram and Darius A. Faroughy and Tjarko Gerdes and Anna Hallin and Gregor Kasieczka and Michael Krämer and Humberto Reyes-Gonzalez and David Shih},
      year={2026},
      eprint={2605.28940},
      archivePrefix={arXiv},
      primaryClass={hep-ph},
      url={https://arxiv.org/abs/2605.28940}, 
}

@inproceedings{julien:hal-00476064,
  TITLE = {{Wasserstein Barycenter and its Application to Texture Mixing}},
  AUTHOR = {Rabin, Julien  and Peyr{\'e}, Gabriel and Delon, Julie and Bernot, Marc},
  URL = {https://hal.science/hal-00476064},
  BOOKTITLE = {{Proc. SSVM'11}},
  ADDRESS = {Israel},
  PUBLISHER = {{Springer}},
  PAGES = {435-446},
  YEAR = {2011},
  HAL_ID = {hal-00476064},
  HAL_VERSION = {v1},
}

@article{bonneel:hal-00881872,
  TITLE = {{Sliced and Radon Wasserstein Barycenters of Measures}},
  AUTHOR = {Bonneel, Nicolas and Rabin, Julien and Peyr{\'e}, Gabriel and Pfister, Hanspeter},
  URL = {https://hal.science/hal-00881872},
  JOURNAL = {{Journal of Mathematical Imaging and Vision}},
  PUBLISHER = {{Springer Verlag}},
  VOLUME = {1},
  NUMBER = {51},
  PAGES = {22-45},
  YEAR = {2015},
  DOI = {10.1007/s10851-014-0506-3},
  HAL_ID = {hal-00881872},
  HAL_VERSION = {v1},
}

@article{DBLP:journals/corr/abs-2001-08361,
  author       = {Jared Kaplan and
                  Sam McCandlish and
                  Tom Henighan and
                  Tom B. Brown and
                  Benjamin Chess and
                  Rewon Child and
                  Scott Gray and
                  Alec Radford and
                  Jeffrey Wu and
                  Dario Amodei},
  title        = {Scaling Laws for Neural Language Models},
  journal      = {CoRR},
  volume       = {abs/2001.08361},
  year         = {2020},
  url          = {https://arxiv.org/abs/2001.08361},
  eprinttype   = {arXiv},
  eprint       = {2001.08361},
  bibsource    = {dblp computer science bibliography, https://dblp.org}
}

@misc{hoffmann2022trainingcomputeoptimallargelanguage,
      title={Training Compute-Optimal Large Language Models}, 
      author={Jordan Hoffmann and Sebastian Borgeaud and Arthur Mensch and Elena Buchatskaya and Trevor Cai and Eliza Rutherford and Diego de Las Casas and Lisa Anne Hendricks and Johannes Welbl and Aidan Clark and Tom Hennigan and Eric Noland and Katie Millican and George van den Driessche and Bogdan Damoc and Aurelia Guy and Simon Osindero and Karen Simonyan and Erich Elsen and Jack W. Rae and Oriol Vinyals and Laurent Sifre},
      year={2022},
      eprint={2203.15556},
      archivePrefix={arXiv},
      primaryClass={cs.CL},
      url={https://arxiv.org/abs/2203.15556}, 
}

@misc{videau2026skalingchinchillasexponentsmeet,
      title={Skaling: Chinchilla's Exponents Meet Kaplan's Coupling}, 
      author={Mathurin Videau and Badr Youbi-Idrissi and David Lopez-Paz and Kartik Ahuja},
      year={2026},
      eprint={2608.07222},
      archivePrefix={arXiv},
      primaryClass={cs.CL},
      url={https://arxiv.org/abs/2608.07222}, 
}

@misc{lovelace2026prescriptivescalinglawsdata,
      title={Prescriptive Scaling Laws for Data Constrained Training}, 
      author={Justin Lovelace and Christian Belardi and Srivatsa Kundurthy and Shriya Sudhakar and Kilian Q. Weinberger},
      year={2026},
      eprint={2605.01640},
      archivePrefix={arXiv},
      primaryClass={cs.LG},
      url={https://arxiv.org/abs/2605.01640}, 
}

@article{HUNGARIAN,
      author         = "Kuhn, Harold. W.",
      title          = "{The Hungarian method for the assignment problem}",
      booktitle      = "{Naval research logistics quarterly}",
      journal        = "Naval research logistics quarterly",
      volume         = "2",
      year           = "1955",
      pages          = "83-97",
      doi            = "",
      SLACcitation   = ""
}

@misc{vaswaniAttentionAllYou2023,
    title = {Attention {Is} {All} {You} {Need}},
    url = {http://arxiv.org/abs/1706.03762},
    doi = {10.48550/arXiv.1706.03762},
    urldate = {2026-09-05},
    publisher = {arXiv},
    author = {Vaswani, Ashish and Shazeer, Noam and Parmar, Niki and Uszkoreit, Jakob and Jones, Llion and Gomez, Aidan N. and Kaiser, Lukasz and Polosukhin, Illia},
    month = aug,
    year = {2023},
    note = {arXiv:1706.03762 [cs.CL]},
}

@article{larkoskiPowerCountingBetter2014,
    title = {Power {Counting} to {Better} {Jet} {Observables}},
    volume = {2014},
    issn = {1029-8479},
    url = {http://arxiv.org/abs/1409.6298},
    doi = {10.1007/JHEP12(2014)009},
    number = {12},
    urldate = {2026-09-26},
    journal = {Journal of High Energy Physics},
    author = {Larkoski, Andrew J. and Moult, Ian and Neill, Duff},
    month = dec,
    year = {2014},
    note = {arXiv:1409.6298 [hep-ph]},
    pages = {9},
}

@misc{kakati2026promptingparticlephysicstokenized,
      title={Prompting Particle Physics: Tokenized Multi-modal Foundation Models for Combinatorially Many Tasks}, 
      author={Nilotpal Kakati and Daniel Murnane and Baran Hashemi and Samuel Klein and Jeffrey Krupa and Eilam Gross and Lukas Heinrich and Michael Kagan},
      year={2026},
      eprint={2609.31862},
      archivePrefix={arXiv},
      primaryClass={hep-ph},
      url={https://arxiv.org/abs/2609.31862}, 
}

\clearpage
\appendix

\section{Methods details}
\label{app:methods}

\subsection{Input representation}
\label{app:inputs}
Every input particle is described by a 12-dimensional vector:
\begin{equation}
  x = \big(\tilde p_T,\ \tilde\eta,\ \sin\phi,\ \cos\phi,\ \tilde v_x,\ \tilde v_y,\ \tilde v_z,\ \mathrm{onehot}_5(\text{class})\big),
\end{equation}
where $\tilde p_T = (\log(1+p_T) - \mu)/\sigma$ and the pseudorapidity and vertex coordinates are normalized with fixed means and standard deviations computed once on the training sample (Table~\ref{tab:var_transform}). The vertex of neutral \pf candidates is set to zero (Section~\ref{sec:methods}).
We also apply a pseudorapidity cut of $|\eta|<2.7$, and during evaluation we restrict the final generated \pf particles with a $|\eta|<2.5$ cut to remain inside the tracker's acceptance region. This allows the generator to model how particles close to the tracker's acceptance region affect the detector response and reconstruction of those inside. 
Particle sequences are $p_T$ sorted and zero-padded to a maximum of $400$ particles and a boolean mask marks valid entries.

\begin{table}[h]
\caption{Fixed feature standardization. $p_T$ is transformed as $\log(1+p_T)$ before standardization; $\phi$ is replaced by $(\sin\phi,\cos\phi)$.}
\label{tab:var_transform}
\centering
\small
\begin{tabular}{lccccc}
 & $\log(1+p_T)$ & $\eta$ & $v_x$ & $v_y$ & $v_z$ \\ \hline
mean $\mu$     & 1.13 & 0.00 & 0.206 & 0.331 & 0.34 \\
std. $\sigma$  & 0.83 & 1.29 & 0.691 & 0.702 & 5.73 \\
\end{tabular}
\end{table}

\subsection{RQ-VAE tokenizer}
\label{app:rqvae}

\paragraph{Architecture}
The tokenizer is a set-to-set autoencoder: both the encoder $E$ and the decoder $G$ act on the $p_T$-ordered  set of particles with bidirectional self-attention.
The encoder and decoder are mirrored and consist of:
\begin{enumerate}
  \item a per-particle input MLP $12 \to 32 \to 128$ (ReLU, with biases);
  \item a $5$-layer transformer at width $d=128$ with $8$ attention heads;
  \item a per-particle output MLP $128 \to 32 \to 16$ (ReLU), producing the latent $z\in\mathbb{R}^{16}$.
\end{enumerate}
Each transformer layer is pre-norm, with affine-free LayerNorm~\citep{ba_layer_2016}. Queries and keys are RMS-normalized per head with a learnable scale (QK-norm, \citealp{dehghaniScalingVisionTransformers2023}).
The feed-forward network is a SwiGLU unit \citep{shazeer_glu_2020} with hidden width $256$ and biases. Attention is computed with the variable-length FlashAttention-2 kernel~\citep{flashattn,dao2023flashattention2fasterattentionbetter} on the unpadded particles of each event. The encoder and decoder parameter count is $0.85$M each, $1.70$M in total.

Codebooks are initialized by $k$-means (100 iterations) on the first training batch \citep{zeghidourSoundStreamEndtoEndNeural2021} and then updated by exponential moving averages of the assigned residuals \citep{oord_neural_2018,razavi_generating_2019} with decay $0.8$. Gradients reach the encoder through the quantizer via the rotation trick \citep{fifty_restructuring_2024}.

\paragraph{Training}
The loss function is the sum of the mean square error (cross-entropy for classes) between the input and its reconstruction and a small per-codebook commitment loss \citep{oord_neural_2018}.
\\The tokenizer is trained on the first $16.5$\,M events of the training dataset. Each event contributes its \truth and \pflow particle sets, for a total of $33$M point-clouds per epoch. We use the AdamW optimizer \citep{loshchilov_decoupled_2019} ($\beta=(0.9,0.999)$, weight decay $0.01$), gradient-norm clipping at $1.0$, and a cosine-decay-schedule learning rate with a brief warmup.
After training, the tokenizer is frozen.

\section{Autoregressive model}
\label{app:ar}

\paragraph{Vocabulary and sequence layout}
The index space contains $V=K+3=8195$ entries, namely a padding token ($0$), the code indices $1,\dots,8192$, a \pf start token and a \tr start token. Code
index $k$ at depth $m$ and the same index at depth $m'\neq m$ map to the same entry, and the depth embedding tells them apart. For an event with $N_\text{TR}$ \truth and $N_\text{PF}$ \pflow particles the input is
\begin{equation}
  \big[\,\texttt{SOS}_\text{TR},\ t_{1,1},\dots,t_{1,4},\ \dots,\ t_{N_\text{TR},4},\ \texttt{pad}\dots\ \big|\
        \texttt{SOS}_\text{PF},\ p_{1,1},\dots,p_{1,4},\ \dots,\ p_{N_\text{PF},4},\ \texttt{pad}\dots\big],
\end{equation}
where each block is padded to $400\times4=1600$ code positions, for a total length of $3202$. Padding never enters computation and does not receive gradients. 

\paragraph{Backbone}
The backbone model is a decoder-only transformer with $L$ layers of width $d_\text{model}$ (Table~\ref{tab:scaling_models}), causal self-attention with $d_\text{head}=64$ (except $d_\text{head}=32$ and $48$ for the two smallest models), and no biases. Each layer follows the HybridNorm arrangement \citep{zhuoHybridNormStableEfficient2025}. Queries, keys and values are each RMS-normalized per head with a learnable scale (QKV-norm) before rotary position embeddings \citep{su_roformer_2023} are applied to queries and keys. Positions are the absolute indices in the padded 3202-long layout, as a result the \pf block always starts at position $1601$, whatever the truth multiplicity.
The feed-forward block is SwiGLU with hidden width $d_\text{ff}=2d_\text{model}$, and a final LayerNorm follows the last layer.

The NTP head is an untied linear map $\mathbb{R}^{d_\text{model}}\to\mathbb{R}^{V}$ with bias. The cardinality head pools the final hidden states over the truth block by a masked mean. The head then applies $\mathrm{Linear}(d_\text{model},d_\text{model})\to\mathrm{GELU}\to\mathrm{Linear}(d_\text{model},401)$, returning logits over $N_\text{PF}\in\{0,\dots,400\}$. Its gradient flows into the shared backbone.

\paragraph{Training}
\label{app:training_ar}
Throughout the scaling study, $N$ follows the convention of \citet{hoffmann2022trainingcomputeoptimallargelanguage} and counts all parameters in the model, including embedding tables, it excludes the frozen tokenizer. A more detailed breakdown of the composition of the models in the study is available in Table~\ref{tab:param_breakdown}

\begin{table}[h]
\caption{Parameter breakdown of the generator models (in millions). $N$ is the total used for fitting.}
\label{tab:param_breakdown}
\centering
\small
\begin{tabular}{lrrrrrr}
Model & Transformer & Code proj. & Token head & Card. head & Embedding tables & $N$ \\ \hline
10M  & 5.244    & 0.004 & 2.106  & 0.169 & 2.305  & 9.827    \\
22M  & 14.747   & 0.007 & 3.155  & 0.302 & 3.457  & 21.668   \\
41M  & 31.460   & 0.009 & 4.204  & 0.468 & 4.609  & 40.750   \\
97M  & 82.578   & 0.013 & 6.302  & 0.899 & 6.914  & 96.706   \\
187M & 167.775  & 0.017 & 8.400  & 1.461 & 9.218  & 186.871  \\
1B   & 1006.638 & 0.035 & 16.792 & 5.018 & 18.436 & 1046.918 \\
\end{tabular}
\end{table}

The loss is applied to the \pf block only: truth-block targets are set to the padding index. During training the countdown embedding uses the true $N_\text{PF}$. All generator runs share one recipe, with no per-size tuning. We employ the AdamW optimizer with $\beta=(0.9,0.999)$, $\epsilon=10^{-5}$ and weight decay $0.01$
applied to all parameters, and a cosine-decay-schedule learning rate with a brief warmup. The decay is tuned to finish at the designed end of the run, and arrives at a tenth of the peak value~\citep{hoffmann2022trainingcomputeoptimallargelanguage}.
The global batch-size is tuned per run to 256 events, using gradient accumulation when needed, with bf16 mixed precision training and gradient-norm clipping at $1.0$.

\paragraph{Inference}

The truth event is tokenized with the frozen encoder and runs through the model once (prefill) to fill the key/value cache.
$N_\text{PF}$ is then sampled from the cardinality head, and exactly $4N_\text{PF}$ codes are sampled autoregressively. Padding and start tokens are masked out from the output's categorical distribution at every step. The sampled codes are summed across depths to $\hat z$ and decoded by the frozen RQ-VAE decoder over the whole generated event.
KV-caching is used at inference to reduce wall-time.
To reduce padding waste, events are first prefilled through the model once, and the cardinality is sampled from the corresponding head.
The events are then reshuffled such that events in the same batch have the same predicted cardinality and thus we avoid wasting decode steps on padding tokens.
A further decoding option, since our cardinality prediction is independent of token sampling (no \texttt{[EOS]} token) is forcing the token cardinality of the target \pflow event corresponding to the fed \truth conditioning, though we find that this \textit{perfect cardinality} method does not yield measurably different results from regularly sampling the cardinality during the prefill phase.
\\Inference-time generation speeds are available in Table~\ref{tab:gen_speed}.

\begin{table}[h]
\caption{Generation throughput of the best run of each model size, measured on $50{,}000$ held-out events. Times include quantization, prefill, autoregressive decode and the RQ-VAE decoding. Each model is tested at the biggest power-of-two batch size $B$ to fit in GPU memory with an $8\%$ margin. The models are tested on NVIDIA A100 SXM4 ($80$\,GB) and NVIDIA H200 ($141$\,GB) GPUs.}
\label{tab:gen_speed}
\begin{center}
\renewcommand{\arraystretch}{1.15}
\setlength{\tabcolsep}{6pt}
\begin{tabular}{lrrrrrr}
& \multicolumn{3}{c}{\textbf{A100}} & \multicolumn{3}{c}{\textbf{H200}} \\
\multicolumn{1}{c}{\textbf{Model}} &
\multicolumn{1}{c}{$\boldsymbol{B}$} & \multicolumn{1}{c}{\textbf{evt/s}} & \multicolumn{1}{c}{\textbf{ms/evt}} &
\multicolumn{1}{c}{$\boldsymbol{B}$} & \multicolumn{1}{c}{\textbf{evt/s}} & \multicolumn{1}{c}{\textbf{ms/evt}}
\\ \hline \\[-6pt]
10M & 1{,}024 & 126.5 & 7.9 & 2{,}048 & 265.1 & 3.8 \\
22M $^\dagger$ & 512 & 15.1 & 66.1 & 1{,}024 & 26.8 & 37.4 \\
41M & 256 & 62.5 & 16.0 & 512 & 132.5 & 7.5 \\
97M & 256 & 38.2 & 26.2 & 256 & 72.7 & 13.8 \\
187M & 128 & 22.0 & 45.4 & 256 & 48.8 & 20.5 \\
1B & 32 & 4.4 & 229.1 & 64 & 12.0 & 83.2 \\
\end{tabular}

\vspace{2pt}
\parbox{0.9\linewidth}{\footnotesize $^\dagger$The 22M model is slower than the parameter count would suggest because of its attention head dimension, $d_\text{head}=48$, for which the inference-running system had no optimized FlashAttn kernel.}

\end{center}
\end{table}

\section{Scaling study methods}
\label{app:scaling}

\subsection{Loss scaling}
\label{app:loss_scaling}
The fitted $\mathcal{L}(N,U,R)$ law uses events as the data indicator: at a fixed test distribution, average tokens per event converge to a constant which the amplitude $B$ absorbs.
Only the size, the data fraction and the epoch count change across the grid. The architecture family, optimizer, peak learning rate, batch size and schedule shape are the same for all runs. 

We fit the four laws of Table~\ref{tab:scaling_loss_fits} directly in the parameter count $N$ and the number of unique events $U$, with the amplitudes $A$, $B$ and $P$ as free parameters. Every law is built from the same two terms:
\begin{align}
  F(N,U,R) &= A\,N^{-\alpha} + B\,[U(1+R)]^{-\beta},\\
  \text{penalty}(N,U,R) &= P\,R^{\delta}\,\big(N/U^{\gamma}\big)^{\kappa}.
\end{align}

Chinchilla is $L_\infty+F$, Skaling is $L_\infty+F^{k}$, Prescriptive is $L_\infty+F+\text{penalty}$, and Custom is $L_\infty+F^{k}+\text{penalty}$.
The penalty is zero at $R=0$. The laws without a repetition term are only defined for single-epoch data and are fitted to the $18$ single-epoch runs. The two repetition-aware laws are fitted to all $48$ runs.
We minimize the unweighted sum of squared residuals in $\mathcal{L}$ with a least-squares solver. The used loss values are the means on the held-out in-distribution testing dataset.
In Table~\ref{tab:scaling_params} we report the fitted parameters, the in-sample Mean Absolute Percentage Error (MAPE) and the coefficient of determination ($R^2$).

\begin{table}[ht]
    \centering
    \caption{Fitted parameters of the loss scaling laws. Chinchilla and Skaling are fitted to the 18 single-epoch runs,
    Prescriptive and Custom to all 48 runs; $N$ and $U$ are in parameters and unique events. Uncertainties are $1\sigma$
    errors over refits with one held-out run.}
    \label{tab:scaling_params}
    \begin{tabular}{lcccc}
 & \textbf{Chinchilla} & \textbf{Skaling} & \textbf{Prescriptive} & \textbf{Custom} \\ \hline \\[-6pt]
$L_\infty$ & $2.67 \pm 0.40$ & $3.191 \pm 0.082$ & $2.80 \pm 0.11$ & $3.255 \pm 0.015$ \\
$A$ & $1300 \pm 4100$ & $34 \pm 31$ & $900 \pm 1200$ & $55 \pm 14$ \\
$\alpha$ & $0.42 \pm 0.18$ & $0.241 \pm 0.052$ & $0.410 \pm 0.092$ & $0.273 \pm 0.017$ \\
$B$ & $(0.9 \pm 2.2)\times10^{4}$ & $106 \pm 93$ & $7400 \pm 3900$ & $114 \pm 24$ \\
$\beta$ & $0.62 \pm 0.15$ & $0.345 \pm 0.070$ & $0.598 \pm 0.036$ & $0.349 \pm 0.016$ \\
$k$ & -- & $2.39 \pm 0.32$ & -- & $2.379 \pm 0.072$ \\
$P$ & -- & -- & $(4.5 \pm 7.9)\times10^{-13}$ & $(3.6 \pm 3.8)\times10^{-11}$ \\
$\kappa$ & -- & -- & $1.260 \pm 0.084$ & $1.411 \pm 0.033$ \\
$\gamma$ & -- & -- & $0.136 \pm 0.041$ & $0.531 \pm 0.041$ \\
$\delta$ & -- & -- & $1.641 \pm 0.092$ & $1.660 \pm 0.028$ \\
\hline
MAPE [\%] & 3.55 & 1.17 & 2.47 & 0.79 \\
$R^2$ & 0.9695 & 0.9961 & 0.9690 & 0.9967 \\
\end{tabular}
\end{table}

\subsection{IsoFLOP analysis}
\label{app:isoflop}

Conventionally, for studies at a fixed compute budget $C$ (introduced by ~\citet{hoffmann2022trainingcomputeoptimallargelanguage}), several model sizes are trained, and their learning-rate schedules end exactly at $C$. Our grid was designed for the $(N,U,R)$ fits instead. Each single-epoch run (repetition is not modeled in these studies) covers one compute value at its endpoint and a range of smaller compute values along its training curve. We therefore build isoFLOP slices from the monitored training curves of the 18 single-epoch runs, following the steps below.
We use the $C \approx 6ND$ \citep{DBLP:journals/corr/abs-2001-08361} approximation, with $D$ being the number of total tokens processed, i.e. $C \approx 6 N [4(n_{\text{TR}} + n_{\text{PF}}) + 2] n_{\text{Evts}}$.
The main difficulty is that for an intermediate checkpoint of a run, its learning rate has not yet fully annealed. Results should be taken with this consideration in mind. The specific retrieval methods are explained in the following section.

Every single-epoch run saves a checkpoint up to every $1/16$ of an epoch, giving 8 to 16 logs per run. Every saved checkpoint's loss is then evaluated on the held out testing dataset.
For a run whose logged range contains the desired compute, we estimate its loss at that compute value by local quadratic interpolation in $\log t$, where $t$ is the optimizer step count: we fit a parabola in $\log t$ through the three logged points closest to $t_w(C)$ (the step at which the run reaches $C$) and evaluate it there.

\begin{figure}[h]
\centering
\includegraphics[width = .7\linewidth]{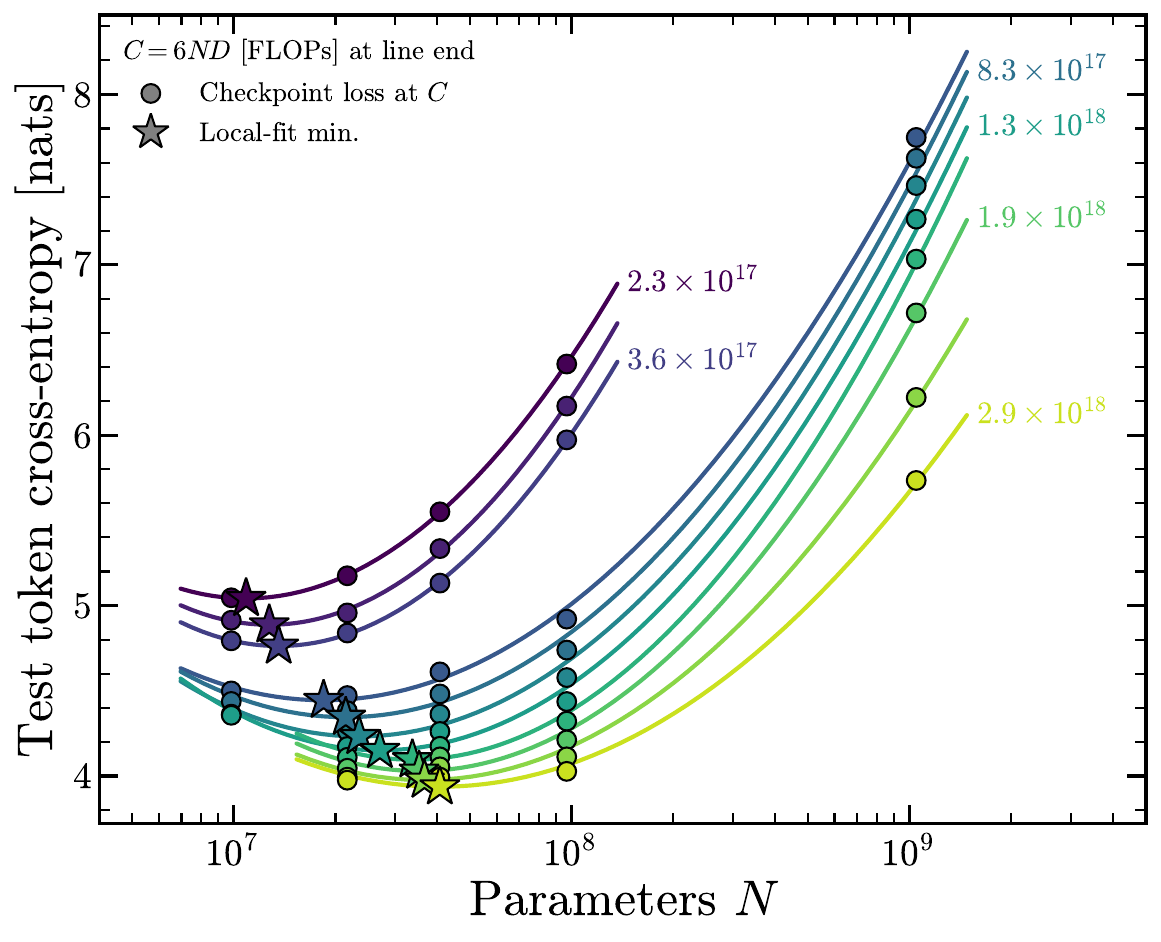}

\caption{IsoFLOP slices fit through the measured loss values. For every slice we show the minima and the compute value.}
\label{fig:isoflop_plots}
\end{figure}

For each $C$ budget, in order to find the minimum achievable loss value we fit a parabola through all the available points, with the criteria that at the specified budget at least 4 different values be available. Fig.~\ref{fig:isoflop_plots} shows the selected isoFLOP slices and associated loss points.

\subsection{Scaling in the physical metrics}
\label{app:physical_scaling}

For every run we generate $50{,}000$ \pflow events conditioned on the same test \truth dataset (temperature $1$, sampled cardinality) and decode it with the frozen tokenizer. 
Before computing any observable, generated and reference \pf events pass the same acceptance, $p_T>1$\,GeV and $|\eta|<2.5$, thresholds.

\paragraph{Residual feature sets.}
We compare distributions of residuals with respect to the truth input rather than raw marginals:
\begin{itemize}
  \item \emph{Event residuals}: for every \fs and \pf event, we compute the difference with respect to the corresponding \tr event of the scalar sum of $p_T$, the missing transverse energies on the $x$ and $y$ axes, and the difference in particle cardinality. 
        
  \item \emph{Particle residuals}: generated particles are Hungarian-matched to the \tr set, with cost $\Delta R$. Pairs with $\Delta R>0.6$ are discarded. Particle continuous features are then considered for the residuals.
  
  \item \emph{Jet-residuals}: anti-$k_T$ jets ($R=0.5$) are clustered at both levels and matched in the same way, with $\Delta R<0.2$. The residuals are those of the jet kinematic and substructure features of Section~\ref{sec:gen-perf}.
\end{itemize}

The full generated manifolds for the different feature sets are subsampled to at most $2\times10^{5}$ rows (this is relevant for the particle-level features).
The computed distance is averaged over $256$ random projections of the one-dimensional $W_1$ between the samples. 
The \emph{noise floor} is the same metric computed between the reference test sample and another statistically identical sample. The RQ-VAE reference is the distance between the reference \pf events and the reconstruction of the second sample from the residual-quantized representation.
For the one-dimensional local metrics we use the exact $W_1$ between residual distributions. 

The loss-metric relation (Eq.~\ref{eq:metric_law}) is fitted by least squares in $\ln\text{SWD}$, only on runs that are
converged and not overfitted. We require (i) $C=6ND\ge3\times10^{17}$, (ii) no overfitting of NTP loss and (iii) no overfitting on the cardinality prediction objective, (iv) more than 2 training tokens per model parameter. 
Intermediate checkpoints are included to probe the loss–metric relationship across training dynamics.

\section{Additional material}
\label{app:plots}

\subsection{Quantization results}
\begin{figure}[h]
\centering
\includegraphics[width = \linewidth]{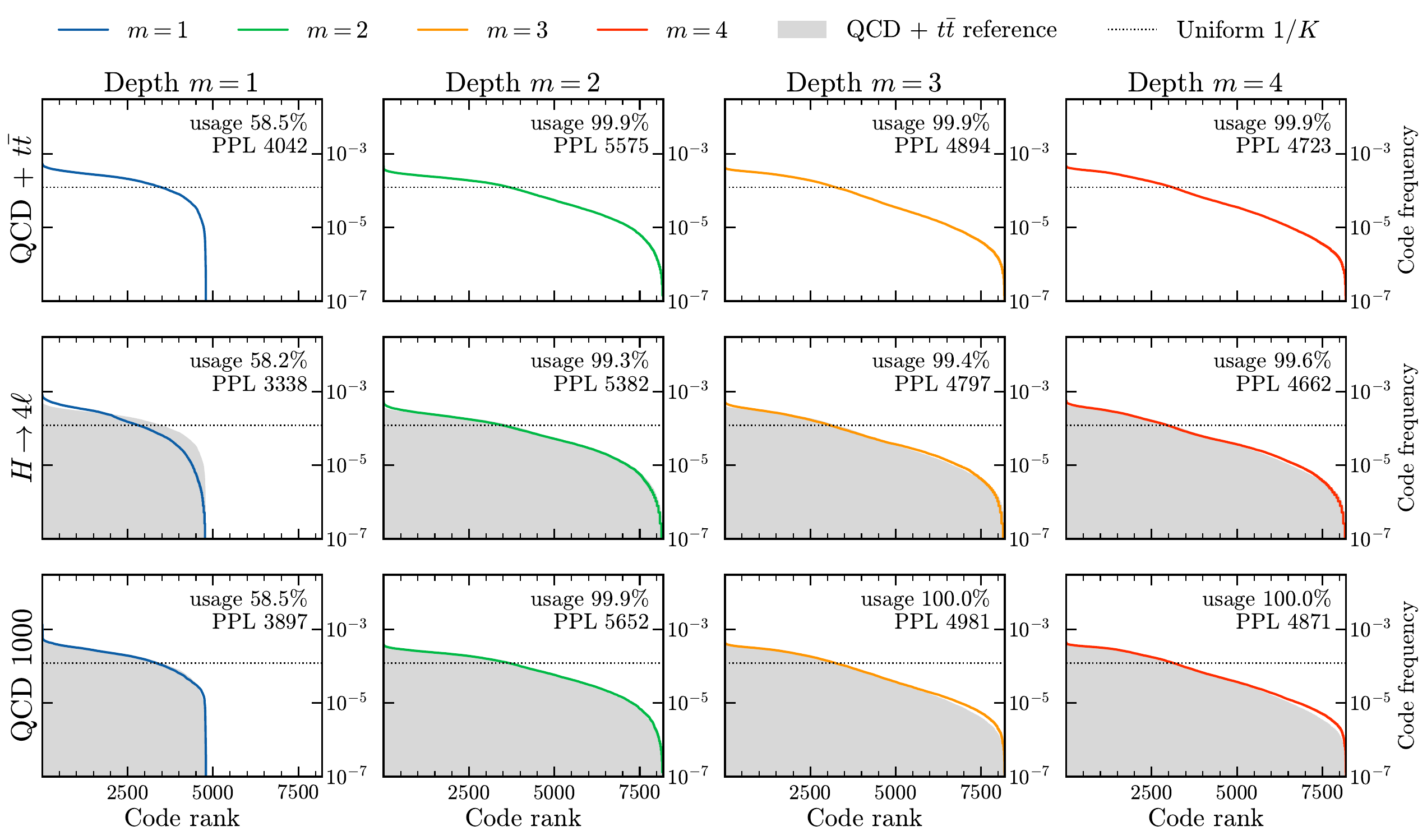} 

\caption{Usage distributions across the testing datasets. Codes are ordered on the $x$ axis according to their usage rank, from most to least used.}
\label{fig:codebook_usage}
\end{figure}

In Figure~\ref{fig:codebook_usage} we report the per-quantizer level codebook usages and perplexities, evaluated across all three testing datasets.

\subsection{Generation results}
\begin{figure}[h]
\centering
\includegraphics[width = \linewidth]{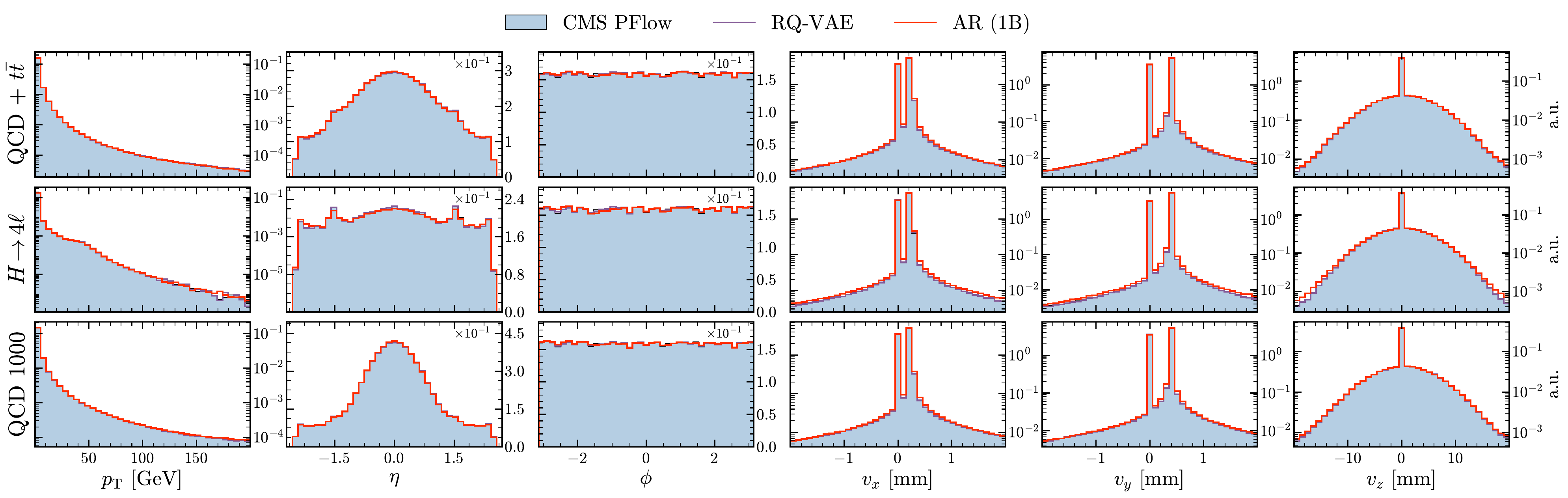} 

\caption{Spectra of particle-level features for all three testing datasets. All histograms are normalized to unity; rows correspond to different datasets/processes while columns correspond to different observables.}
\label{fig:part_marginals}
\end{figure}

\begin{figure}[h]
\centering
\includegraphics[width = \linewidth]{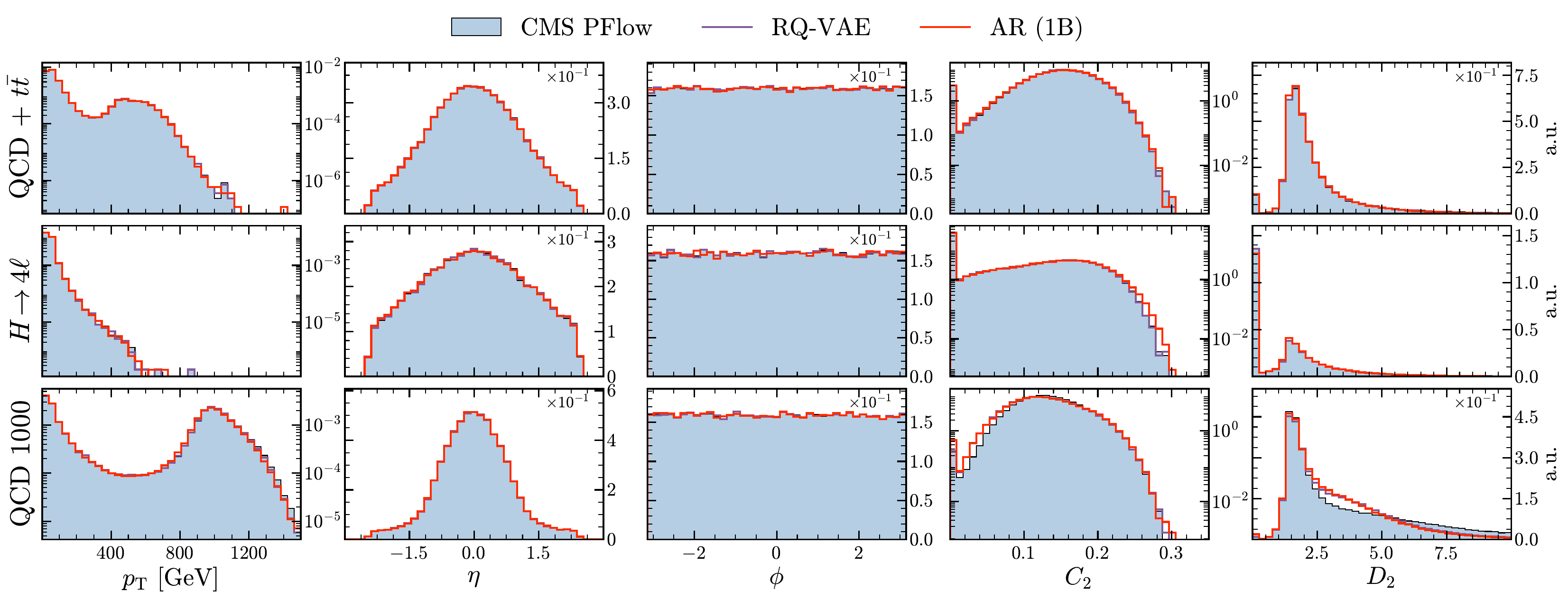} 

\caption{Spectra of jet-level features for all three testing datasets. All histograms are normalized to unity; rows correspond to different datasets/processes while columns correspond to different observables.}
\label{fig:jet_marginals}
\end{figure}

\begin{figure}[h]
\centering
\includegraphics[width = \linewidth]{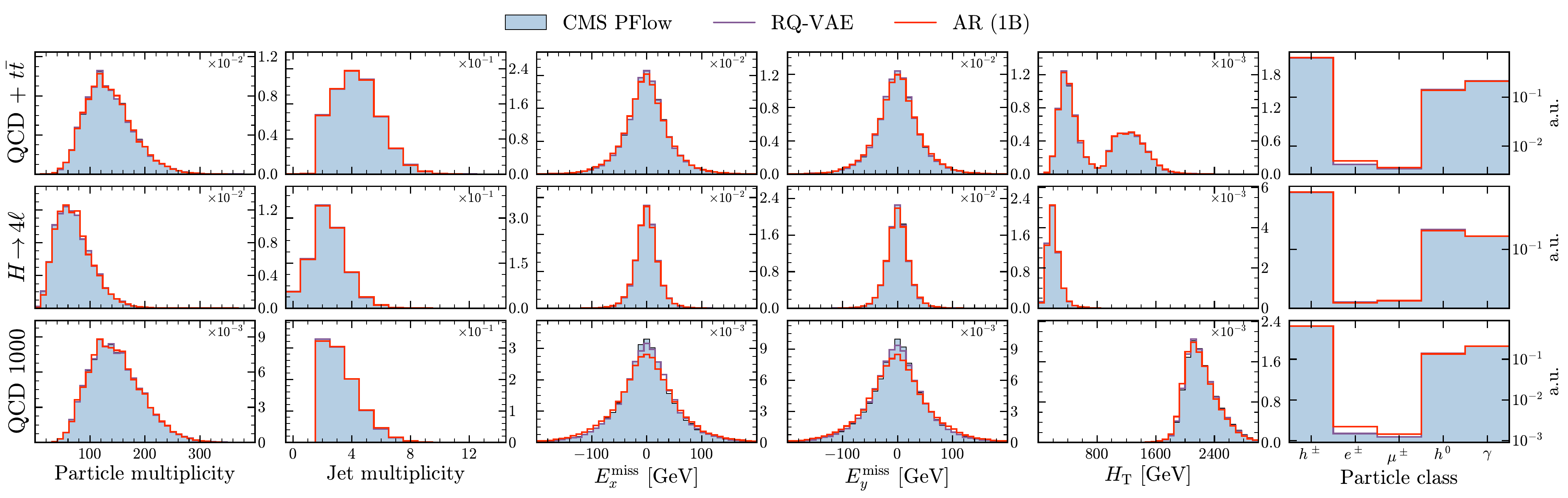} 

\caption{Spectra of event-level features for all three testing datasets. All histograms are normalized to unity; rows correspond to different datasets/processes while columns correspond to different observables.}
\label{fig:event_marginals}
\end{figure}

We present the marginal feature distribution corresponding to the residual distributions presented in Section~\ref{sec:gen-perf}. The spectra of particle features are available in Figure~\ref{fig:part_marginals}, those of jets in Figure~\ref{fig:jet_marginals}, and finally the distributions of event-level features are shown in Figure~\ref{fig:event_marginals}.
A comparison of the generation performance of different ablations is available in Table~\ref{tab:w1_models}

\begin{table}[h]
\caption{Per-feature $W_1$ distance between the \fs and \pf residual distributions, in the units of each feature, for the models trained on the full dataset for $4$ epochs, on the held-out in-distribution test set ($50{,}000$ events). The RQ-VAE column is the tokenizer reconstruction of an independent $50{,}000$-event sample with the same process mix, and the noise column the distance between two such independent \pf samples. The best model in each row is in bold.}
\label{tab:w1_models}
\begin{center}
\small
\renewcommand{\arraystretch}{1.15}
\setlength{\tabcolsep}{3.5pt}
\begin{tabular}{llcccccccc}
\multicolumn{1}{c}{\bf Residual} & \multicolumn{1}{c}{\bf Unit} &
\multicolumn{1}{c}{\bf 10M} & \multicolumn{1}{c}{\bf 22M} & \multicolumn{1}{c}{\bf 41M} &
\multicolumn{1}{c}{\bf 97M} & \multicolumn{1}{c}{\bf 187M} & \multicolumn{1}{c}{\bf 1B} &
\multicolumn{1}{c}{\bf RQ-VAE} & \multicolumn{1}{c}{\bf Noise}
\\ \hline \\[-6pt]
\multicolumn{10}{l}{\emph{Event residuals}} \\
$\Delta H_T/H_T$             & --   & 0.0212 & 0.0154 & 0.0102 & 0.00538 & 0.00526 & \textbf{0.00262} & 0.00144 & 0.00132 \\
$\Delta E_x^{\mathrm{miss}}$ & GeV  & 17.3 & 9.98 & 6.69 & 3.81 & 2.97 & \textbf{1.81} & 0.363 & 0.271 \\
$\Delta E_y^{\mathrm{miss}}$ & GeV  & 17.1 & 10.3 & 6.82 & 3.77 & 3.09 & \textbf{1.79} & 0.251 & 0.347 \\
$\Delta E_T^{\mathrm{miss}}$ & GeV  & 22.9 & 13.2 & 8.60 & 4.62 & 3.78 & \textbf{2.09} & 0.439 & 0.491 \\
$\Delta N/N$                 & --   & 0.0195 & 0.0131 & 0.00940 & 0.0130 & 0.00629 & \textbf{0.00513} & 0.00793 & 0.00509 \\
$\Delta N_{\mathrm{jets}}$   & --   & 0.232 & 0.144 & 0.101 & 0.0646 & 0.0541 & \textbf{0.0292} & 0.00656 & 0.00612 \\
[0.4em]
\multicolumn{10}{l}{\emph{Jet residuals}} \\
$\Delta p_T/p_T$             & --   & 0.0708 & 0.0475 & 0.0344 & 0.0212 & 0.0171 & \textbf{0.00990} & 0.000811 & 0.000869 \\
$\Delta\eta$                 & --   & 0.0306 & 0.0208 & 0.0152 & 0.0112 & 0.00994 & \textbf{0.00787} & 0.00296 & 0.000114 \\
$\Delta\phi$                 & rad  & 0.0117 & 0.00783 & 0.00609 & 0.00424 & 0.00364 & \textbf{0.00248} & 0.000470 & 0.000105 \\
$\Delta C_2$                 & --   & 0.00792 & 0.00504 & 0.00382 & 0.00237 & 0.00207 & \textbf{0.00127} & 0.000612 & 0.000198 \\
$\Delta D_2$                 & --   & 0.433 & 0.327 & 0.271 & 0.240 & 0.218 & \textbf{0.189} & 0.135 & 0.0112 \\
[0.4em]
\multicolumn{10}{l}{\emph{Particle residuals}} \\
$\Delta p_T/p_T$             & --   & 1.12 & 0.813 & 0.670 & 0.526 & 0.477 & \textbf{0.392} & 0.212 & 0.00564 \\
$\Delta\eta$                 & --   & 0.0285 & 0.0189 & 0.0154 & 0.0113 & 0.0103 & \textbf{0.00809} & 0.00357 & 0.000163 \\
$\Delta\phi$                 & rad  & 0.0185 & 0.0111 & 0.00946 & 0.00625 & 0.00555 & \textbf{0.00398} & 0.00183 & 0.000225 \\
$\Delta v_z$                 & mm   & 0.375 & 0.232 & 0.245 & \textbf{0.160} & 0.229 & 0.187 & 0.0356 & 0.0138 \\
$\Delta v_x$                 & mm   & \textbf{0.0292} & 0.0414 & 0.0461 & 0.0328 & 0.0449 & 0.0309 & 0.00145 & 0.00143 \\
$\Delta v_y$                 & mm   & 0.0310 & 0.0429 & 0.0513 & 0.0349 & 0.0458 & \textbf{0.0287} & 0.00178 & 0.00250 \\

\end{tabular}
\end{center}
\end{table}

\end{document}